\documentclass[a4paper,twocolumn,showpacs,
]{revtex4}

\usepackage{psfrag}
\usepackage{graphicx}
\usepackage{amssymb}
\usepackage{amsmath,amsfonts,latexsym}
\usepackage{dcolumn}               

\DeclareMathOperator{\tr}{tr}

\DeclareMathOperator{\str}{str}

\DeclareMathOperator{\diag}{diag}

\newcommand{\vect}[1]{{\mathbf #1}}
\newcommand{\vectgr}[1]{{\boldsymbol#1}}    

\newcommand{\Frac}[2]{\displaystyle\frac{#1}{#2}}
\newcommand{\smc}[1]{\text{\sc{#1}}}

\begin{document}


\title{Universality of Parametric Spectral Correlations: Local {\em versus}
       Extended Perturbing Potentials}

\author{F. M. Marchetti}

\author{I. E. Smolyarenko}

\author{B. D. Simons}

\affiliation{Cavendish Laboratory, Madingley Road, Cambridge CB3 \
             0HE, UK}

\date{April 30, 2003}       

\begin{abstract}
We explore the influence of an arbitrary external potential
perturbation $V$ on the spectral properties of a weakly disordered
conductor. In the framework of a statistical field theory of a
nonlinear $\sigma$-model type we find, depending on the range and the
profile of the external perturbation, two qualitatively different
universal regimes of parametric spectral statistics
(i.e. cross-correlations between the spectra of Hamiltonians $H$ and
$H+V$). We identify the translational invariance of the correlations
in the space of Hamiltonians as the key indicator of universality, and
find the connection between the coordinate system in this space which
makes the translational invariance manifest, and the physically
measurable properties of the system. In particular, in the case of
localized perturbations, the latter turn out to be the eigenphases of
the scattering matrix for scattering off the perturbing potential
$V$. They also have a purely statistical interpretation in terms of
the moments of the level velocity distribution. Finally, on the basis
of this analysis, a set of results obtained recently by the authors
using random matrix theory methods is shown to be applicable to a much
wider class of disordered and chaotic structures.
\end{abstract}








\maketitle

\section{Introduction}
\label{sec:intro}
The statistical approach to the study of the quantum spectra of
complex Hamiltonians $H$ was pioneered in the 1950's in connection
with the study of resonances in nuclear
scattering~\cite{wigner,dyson_a}. The approach was formalized by
Dyson~\cite{dyson_a}, who identified three universal types of spectral
statistics as determined by the symmetry properties of the
corresponding ensembles of random Hamiltonians. The theory of Wigner
and Dyson, together with its extensions, is usually referred to as
random matrix theory (RMT). Within RMT, spectral correlations of each
type (symmetry class) of statistics are described by the corresponding
set of universal functions, once the energies are rescaled by the
average level spacing $\bar{\Delta}$. The latter is the \emph{only}
non-universal parameter of the theory, where non-universality implies
dependence on the specifics (other than the symmetry class) of the
distribution functions defining the Hamiltonian ensemble. The symmetry
classes are labeled by the index $\beta$ which takes the values
$\beta=1$ for real (time-invariant) Hamiltonians (so-called orthogonal
class), $\beta=2$ for complex Hamiltonians (unitary class), and
$\beta=4$ for Hamiltonians with broken spin-rotation symmetry
(symplectic class).

For example, in the unitary class ($\beta=2$), the normalized
two-point correlation function of the density of states (DoS), $\nu
(\varepsilon) = \tr \delta (\varepsilon - H)$,
\begin{equation}
 R_2 (\Omega) = \bar{\Delta}^2\langle \nu (\varepsilon + \Omega/2) \nu
  (\varepsilon - \Omega/2)\rangle 
\label{eq:ancor}
\end{equation}
can be expressed as
\begin{multline}
  R_2 (s) = 1 + \Frac{1}{2} \mathrm{Re} \int_{1}^{\infty} d\lambda_1
  \int_{-1}^{1} d\lambda\; e^{i \pi s_{+} (\lambda_1 - \lambda)} \\
  = \delta(s) + 1 - k^2 (s)\; ,
\label{eq:rmtbe}
\end{multline}
where
\begin{equation}
  k (s) = \Frac{\sin (\pi s)}{\pi s}\; ,
\label{eq:kfunc}
\end{equation}
$s = \Omega/\bar{\Delta}$, and where $s_{+}=s +i0$. The delta function
in~\eqref{eq:rmtbe} represents the autocorrelation of an energy level
with itself; the constant term is the disconnected contribution, and
the oscillating term corresponds to non-trivial correlations between
different levels. Similar, although slightly more convoluted,
expressions are well known for other symmetry classes.

As became clear in the subsequent development of the theory,
Wigner-Dyson statistics possesses a remarkable universality: beyond
the manifestly model constructions of RMT, Wigner-Dyson statistics
occurs both in quantum weakly disordered
systems~\cite{efetov_p,efetov} $H = H_0 + H_{\text{dis}}$, where the
statistical description follows naturally from the ensemble of
realizations of the disordered part of the Hamiltonian
$H_{\text{dis}}$, as well as in \emph{individual} classically chaotic
Hamiltonians~\cite{bohigas_giannoni} upon averaging over a
sufficiently wide spectral window. The status of the last two
statements is somewhat different: while the former has been rigorously
established~\cite{efetov_p,efetov}, the latter statement is known as
the Bohigas-Giannoni-Schmit (BGS)~\cite{bohigas_giannoni} {\em conjecture}.
Crucially, the
existence of a statistical ensemble allows average properties of the
disordered Hamiltonian to be formulated in the framework of a quantum
field theory of nonlinear sigma model type (NL$\sigma$M) valid in the
limit $g\gg 1$.
Here, the dimensionless conductance $g$ denotes a
non-universal parameter which depends on the deterministic part of the
Hamiltonian $H_0$ -- and through it on the sample geometry -- as well
as on details of the distribution of $H_{\text{dis}}$. By contrast,
properties of an individual system are much less amenable to treatment
by means of a statistical field theory and, despite recent encouraging
progress~\cite{andreev_agam_simons}, a convincing formal proof of the
BGS conjecture is lacking. The existence of a proof of Wigner-Dyson
statistics in quantum disordered systems generates a dichotomy of
approaches: if universality is \emph{assumed} or established by other
means, RMT calculations can be employed to obtain specific answers. On
the other hand, NL$\sigma$M calculations often represent the only
available route to \emph{establish} universality and to explore
deviations from it (typically as expansions in $1/g$).

A corollary of the above universality is the fact that, apart form a
possible modification of the average DoS $\bar{\Delta}^{-1}$, the
spectral statistics of $H=H_0+H_{\text{dis}}$ and
$H_{V}=H_0+H_{\text{dis}}+V$ are indistinguishable \cite{zee} unless
$V$ is so large that $H_0+H_{\text{dis}}$ generate only perturbative
corrections to its eigenvalues (the same condition, of course, applies
to the relation between $H_0$ and $H_{\text{dis}}$). Such universality
dictates that the statistics of spectral lines tracing the evolution
of eigenvalues as functions of a coordinate $X$ along a typical curve
in the Hamiltonian space connecting $H_0$ to $H_0+V$ must possess some
kind of stationarity \cite{dyson_b}. This stationarity is reflected,
in turn, in the universality of parametric spectral correlations
\cite{parametric} (i.e. cross-correlations between the spectra of $H$
and $H_{V}$ averaged over the realizations of
$H_{\text{dis}}$). Indeed, as first pointed out in Refs.
\cite{simons_altshuler,szafer,altshuler_simons},
for a broad class of $V$'s the parametric spectral correlators are
expressed via another set of universal functions (again determined
only by the symmetry of the Hamiltonian ensemble) which depend, in
addition to the level density, on a \emph{single} extra non-universal
parameter -- the dispersion $C(0)$ of the distribution of level
velocities $\bar{\Delta}^{-1} \partial \varepsilon_i/\partial X \equiv
\bar{\Delta}^{-1} \partial_{X} \varepsilon_i$, where $\varepsilon_i$
is a typical eigenlevel of $H$. (The use of the notation $C(0)$ for
the dispersion of the distribution of level velocities is explained by
the fact that it is the limiting value of the level velocity
correlation function $C(X) = \bar{\Delta}^{-2} \langle
\partial_{\bar{X}} \varepsilon_i (\bar{X}) \partial_{\bar{X}}
\varepsilon_i (\bar{X} + X)\rangle$
\cite{simons_altshuler,szafer,altshuler_simons}.)
If the curve is a `straight line' $H+XV$, the rescaled parameter
\begin{equation}
  x_0 = X \sqrt{C(0)}
\label{eq:xdefi}
\end{equation}
is an effective measure of the strength of $XV$. Eq.~\eqref{eq:xdefi}
is, in essence, a relation between two phenomenological parameters,
$x_0$ and $C(0)$. It is supplemented by a `microscopic' definition of
$x_0$ in terms of
$XV$~\cite{simons_altshuler,altshuler_simons} 
\begin{equation}
  x_0^2 = \Frac{8}{\bar{\Delta}^2} X^2 \langle\langle V \Pi V
  \rangle\rangle\; ,
\label{eq:xexpl}
\end{equation}
where the double angle brackets denote the matrix element of $\Pi$ on
the vector space spanned by $V$, and the definition of the
(non-universal) operator $\Pi$ strongly depends on $H_0$ and the
distribution of $H_\text{dis}$. In a disordered metal, $\Pi$ becomes a
resolvent of the diffusion
operator~\cite{simons_altshuler,altshuler_simons}, and, if $V$ is
diagonal in the coordinate representation, the matrix element
simplifies to $\langle\langle V \Pi V\rangle\rangle = L^{-2d} \int
d\vect{r} d\vect{r}' V(\vect{r})\Pi(\vect{r}, \vect{r}')V(\vect{r}')$,
where $\Pi(\vect{r}, \vect{r}')$ is the matrix element of $\Pi$ in the
coordinate basis, $L$ is the linear size of the sample, and $d$ are
the space dimensions. The parameter $x_0$ can be thought of as a
scalar norm (in the space of Hamiltonians) of the `distance' between
$H$ and $H + XV$, and Eq.~\eqref{eq:xexpl} can correspondingly be
viewed as fixing a particular prescription for the norm.

The importance of Eq.~\eqref{eq:xdefi} is underscored by the fact that
in practice, measuring $V$ directly is often impossible or very
difficult, and the Green function of the diffusion operator is itself
a complicated object in an irregularly shaped sample. Consequently,
direct evaluation of~\eqref{eq:xexpl} is seldom feasible. On the other
hand, collecting statistics on level velocities is relatively
straightforward, so that Eq.~\eqref{eq:xdefi} is the \emph{only} link
between the formal parameter $x_0$ and an experimentally measurable
quantity.

[A brief remark on notation: we denote the parametric perturbation as
$XV$ whenever the emphasis is on the continuous evolution of the
spectrum of $H + XV$ as a function of $X$ along a `straight line' $0
\leq X \leq 1$. In what follows non-linear parameter dependence of $V$
will become important, so it will be convenient to revert to denoting
the perturbation as $V$ with its parameter dependence implicit from
the context.]

The `broad class of $V$'s' which was referred to above is
distinguished by the following properties. First of all, the main
feature of the class of $V$'s to which the theory of
Refs.~\cite{simons_altshuler,altshuler_simons}
applies is that the level velocity distribution is Gaussian, and is
thus fully characterized by its second cumulant. It is, of course,
generic in some sense, as the Gaussianity of the level velocity
distribution is the result of the central limit theorem which comes
into force for any $V$ which is \emph{global}, or \emph{extended},
i.e. when $V$ is a generic `full' matrix in the Hilbert space defined
by a typical realization of $H_0 + H_{\text{dis}}$ (a more formal
definition is $\tr V^4/(\tr V^2)^2\to 0$ in the thermodynamic
limit). In application to disordered metals, such perturbations are
easily realized as, e. g., inhomogeneous electric or magnetic fields
acting on the whole, or a substantial part of, the sample volume.

Secondly, the fact that $x_0$ is linear in $X$ implies that $V$ is, in
some sense, small. More precisely, individual matrix elements of $V$
are small as compared to the mean level spacing $\bar{\Delta}$, even as
the overall effect of $V$ is finite due to its global extent. It
cannot be too strongly emphasized that, for a global perturbation, the
requirement that its matrix elements are small is not an artificial
constraint, but a self-consistent condition for the existence of
finite correlations between the spectra of $H$ and $H + V$: if this
condition is violated, the spectrum of $H + V$ is so `scrambled' with
respect to the spectrum of $H$ that the residual correlations vanish
as inverse powers of the system volume.

There exists, however, a class of perturbations $V$ which don't quite
fall into the paradigm of
Refs.~\cite{simons_altshuler,altshuler_simons}. In many practical
applications, for example, $V$ cannot be characterized as extended. A
bistable defect jumping between two nearby configurations, a local STM
tip, defects created under irradiation, etc., belong to a very
different type of a perturbing Hamiltonian. On an intuitive level they
are easily seen to be `local', and formally $\tr V^4/(\tr V^2)^2$ is
finite for such perturbations. Even more crucially, the corresponding
distribution of level velocities is not Gaussian [e.g., it is
Poissonian for a moving local defect in a disordered metal for
$\beta=2$, see Eq.~\eqref{eq:defve} below], so the results of
Refs.~\cite{simons_altshuler,altshuler_simons} need to be modified.

\begin{figure}
\begin{center}
\includegraphics[width=1\linewidth,angle=0]{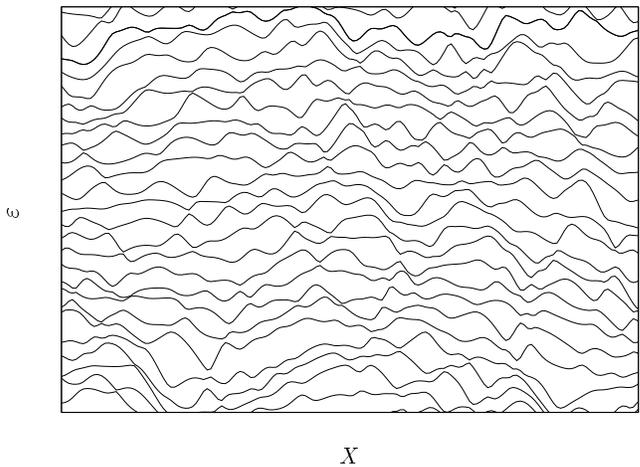}
\end{center}
\caption{\small 
Parametric dependence of the
energy levels of a tight-binding Anderson
Hamiltonian on a 
$20 \times 20$ lattice 
as a function of the strength $X$ of a fixed global
perturbation (arbitrary units). To place the system in the unitary
symmetry class, we have imposed an external random vector potential in
addition to the random scalar potential.}
\label{fig:rankn}
\end{figure}

\begin{figure}
\begin{center}
\includegraphics[width=1\linewidth,angle=0]{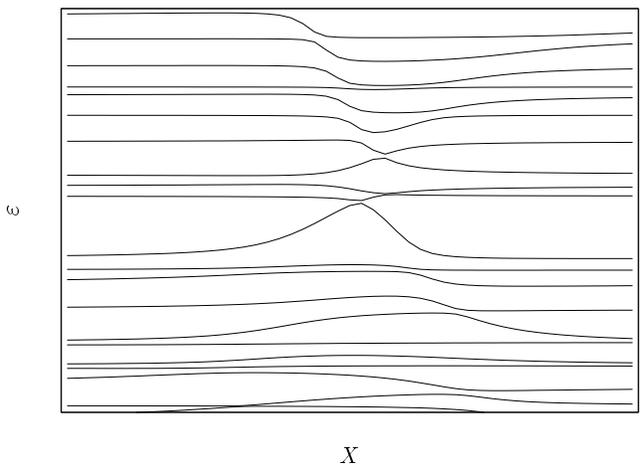}
\end{center}
\caption{\small 
Parametric dependence of the energy levels of a
time-reversal-invariant tight-binding Anderson Hamiltonian on a $20
\times 20$ lattice as a function of the overall strength $X$ of an
external perturbation localized on two fixed sites (arbitrary units).}
\label{fig:levr2}
\end{figure}

Parametric spectral correlations induced by non-exten\-ded perturbing
potentials have been studied previously in the RMT framework
\cite{SMS,SS1,SS2}, leading to a comprehensive set of results for a
large class of parametric correlation functions in the unitary
ensemble. As noted above, the use of RMT presupposes universality. At
the same time, matrix elements of a local perturbation need not be
small. As a typical example, let us consider the potential arising in
the x-ray emission/absorption spectra in disordered metals, and the
associated issue of Anderson orthogonality
catastrophe~\cite{anderson}. The potential $U(\vect{r})$ of a charged
ion created after an electron has been knocked out of an inner shell
by a photon may typically have a magnitude comparable to the Fermi
energy $\varepsilon_F$ and a spatial extent of the order of the Fermi
wavelength $\lambda_F = 2\pi/p_F$. For comparison, an example of a
global potential to which analysis of
Refs.~\cite{simons_altshuler,altshuler_simons} fully applies is
external magnetic field generating a unit of flux through the sample
area. The corresponding effective potential is easily estimated as
$(e/mc) \vect{p}\vect{A} \sim \varepsilon_F/p_F L$, where $\mathbf{A}$
is the vector potential. In the former case, the potential, although
spatially localized, is larger by a factor of square root of the
sample volume (in $d=2$ dimensions). The qualitative difference
between the two types of perturbations is vividly illustrated in
Figs.~\ref{fig:rankn} and~\ref{fig:levr2} which depict the evolution
of a set of levels of a typical realization of $H_0 +
H_{\text{dis}}+XV$ as a function of $X$ for the cases of global and
local perturbations, respectively. Note, for example, that in the case
of a local perturbation, even very large values of $X$ do not lead to
a complete `scrambling' of the spectrum.


Given these qualitative differences, the universality of the
corresponding parametric correlations cannot be automatically deduced
from the analysis of
Refs.~\cite{simons_altshuler,altshuler_simons}. In this paper, we
utilize the NL$\sigma$M approach to prove the universality of the
results in~\cite{SMS,SS1,SS2} and to establish their limits of
applicability and, crucially, their parametrization in terms of
non-universal ($H_0$-dependent) quantities. We also establish a
precise criterion for the crossover between the localized
\cite{SMS,SS2} and extended \cite{simons_altshuler,altshuler_simons}
regimes. The NL$\sigma$M also affords an extension of some of the
previously obtained results to the orthogonal ensemble, although a
complete counterpart of the results of Refs. \cite{SMS,SS1,SS2} in the
orthogonal case is unattainable at present.

\section{Parametric spectral correlation functions}
\label{sec:param}
The parametric analog of Eq.~\eqref{eq:ancor} -- and the basic object
to which we devote most of the attention in this paper -- is the
cross-correlation function
\begin{equation}
  R_{11}(\Omega,V) = \bar{\Delta}^2 \langle \nu (\varepsilon +
  \Omega/2 ; 0) \; \nu (\varepsilon - \Omega/2 ; V)\rangle \; ,
\label{eq:r11de}
\end{equation}
where $\nu(\varepsilon ; V) = \tr \delta (\varepsilon - H - V)$.  The
notation is dictated by the fact that $R_{11}$ is the special case of
the multi-point correlation function $R_{nm}$ involving $n$ densities
$\nu (\varepsilon_i;0)$ and $m$ `shifted' densities
$\nu(\varepsilon_j;V)$. (Within this classification scheme, $R_2$ is
more properly denoted as $R_{20}$.)  Its universal form (in the case
of random Hamiltonians of unitary symmetry) is given, as established
in Refs.~\cite{simons_altshuler,altshuler_simons}
for global $V$, by the following expression:
\begin{multline}
  R_{11} (s , V) - 1 \\ 
  = \Frac{1}{2} \mathrm{Re} \int_1^\infty d\lambda_1 \int_{-1}^1 d\lambda\;
  e^{i \pi s_+ (\lambda_1 - \lambda) - \sigma(\lambda,\lambda_1;x_0)} 
  \; ,
\label{eq:sline}
\end{multline}
where
\begin{equation}
  \sigma (\lambda,\lambda_1;x_0) \equiv \sigma_{\text{gl}}
  (\lambda,\lambda_1;x_0) = \pi^2 x_0^2 (\lambda_1^2-\lambda^2)/2 \; ,
\label{eq:sigma}
\end{equation}
and $x_0$ is given by Eq.~\eqref{eq:xexpl}. Note that the resolvent of
the diffusion operator $\Pi(\vect{r},\vect{r}')$ is a long-range
object, so that the extended character of $V$ is essential in the
structure of Eq.~\eqref{eq:xexpl}. Note also that $\Pi$ is smooth on
the scale of the mean free path $\ell$, suppressing the contribution
of the fast components of $V$ to~\eqref{eq:xexpl}.

The basic conclusion of the analysis presented in this paper is that
in the case of a local perturbation (or a perturbation which has both
global and local components) the overall structure of
Eq.~\eqref{eq:sline} is preserved, while the function $\sigma
(\lambda,\lambda_1;x_0)$ is generalized to
\begin{gather}
\nonumber
   \sigma (\lambda,\lambda_1;x_0,\vect{x}) = \sigma_{\text{gl}}
   (\lambda,\lambda_1;x_0) + \sigma_{\text{loc}}
   (\lambda,\lambda_1;\vect{x})\\ 
   \sigma_{\text{loc}}(\lambda,\lambda_1;\vect{x})  = \sum_{a=1}^r
   \left[\ln\left(1 + ix_a \lambda_1\right) -\ln \left(1 + ix_a
   \lambda\right)\right] \; ,
\label{eq:sgene}
\end{gather}
where the `local' part $\sigma_{\text{loc}}$ of $\sigma$ depends on a vector of
parameters $\vect{x}$. The vector $\vect{x}$ has a finite number $r$
of components, reflecting the finite extent of the local component of
$V$. We also find an additional contribution to $x_0^2$, which
properly accounts for the fast spatial oscillations in the extended
parts of $V$,
\begin{equation}
  \delta x_0^2 = 4 \bar{\nu}^2 \int d\vect{r} d\vect{r}'
  V(\vect{r}) f_d^2 (\vect{r} - \vect{r}') V(\vect{r}') X^2 \; ,
\label{eq:addx0}
\end{equation}
where $\bar{\nu} = 1/\bar{\Delta} L^d$ is the average local DoS in a
$d$-dimensional sample. The Friedel function $f_d
(\vect{r})$ represents the averaged Green function of a disordered
system, and is given by
\begin{equation}
  f_d (\vect{r}) = J_{d/2 - 1} (p_Fr) e^{-r/2\ell}\; ,
\label{eq:ffrie}
\end{equation}
where $J_\eta$ is the Bessel function of $\eta$-th order.

In what sense is Eq.~\eqref{eq:sgene} universal, depending, as it
does, on a whole list of parameters? To answer this question one needs
to clarify whether analogs of Eq.~\eqref{eq:xdefi} can be defined,
i.e., whether these correlation functions are determined by an (at
most finite) set of phenomenological parameters rather than the
thermodynamically large number of variables in the pair of matrices
$H_0$ and $V$.

As will be explained shortly, $x_a$ are nonlinear functions of
$XV$. In other words, the transformation analogous
to~\eqref{eq:xdefi}, if it exists at all, is \emph{non-linear} -- in
contrast to the simple rescaling required in the global case. The
nonlinearity means that, unlike the case of a global perturbation, the
overall magnitude of the perturbation $X$ is no longer a `natural'
variable with respect to which level velocity should be defined. It is
not even {\em a priori} clear that such a `natural' variable exists.

There is, however, an extra requirement that can be imposed to make
the parametrization unique. To help motivate it, let us note that, in
the global case, $\sigma$ as a function of $x_0$ is essentially a
Taylor expansion where terms of the order higher than $2$ vanish in
the thermodynamic limit. As a result, correlation functions between $H
+ X_1V$ and $H + X_2V$ for arbitrary $X_1$ and $X_2$ are functions of
the (properly rescaled) difference $X_2 - X_1$. This translationally
invariant parametrization is a direct consequence -- in fact, the
primary manifestation -- of the principle of stationarity discussed in
the Introduction. On general grounds, such stationarity should
characterize parametric correlations induced by local perturbations as
well. Thus, there must exist a `natural' set of variables
$\vect{y}(\vect{x})$ possessing the following property: there is at
least one curve connecting $H_0$ and $H_0 + V$ such that the
correlations between any two Hamiltonians on this curve corresponding
to two different values $X_1$ and $X_2$ are functions of the
differences $\vect{y}(\vect{x}_2) - \vect{y}(\vect{x}_1)$.  The
distribution of level velocities along this curve is independent of
the position on the curve. In fact, for $r>1$, such curves are not
unique, but rather span an $r$-dimensional manifold.

It is worth emphasizing that the existence of such parametrization is
a non-trivial property even for a single parameter ($r=1$). For
example, let us consider an arbitrary function $f(X_1 , X_2)$. The
existence of a `natural' set of variables is equivalent to the
requirement that for \emph{any} $X_1$ and $X_2$ the function $f(X_1 ,
X_2)$ can be represented as $g(h(X_2) - h(X_1))$ for some functions
$g$ and $h$. Such a representation clearly exists only for a very
restricted subset of all possible $f$. As follows from the analysis
below, the `natural' parametrization does indeed exist, and it is
effected by the substitution
\begin{equation}
  y_a = \arctan x_a \; .
\label{eq:arctn}
\end{equation}
As will be shown in Section~\ref{sec:ftheo}, $x_a$ are the eigenvalues of the
\emph{reactance matrix}
\begin{equation}
  \mathcal{R} \equiv i (\mathcal{S} - \openone) (\mathcal{S} +
  \openone)^{-1}\; ,
\end{equation}
where the scattering matrix $\mathcal{S}$ describes scattering off $V$
with free propagation controlled by the Green function of $H = H_0 +
H_{\text{dis}}$ averaged over $H_{\text{dis}}$. The emergence of the
reactance matrix as a key object governing parametric correlations is
rather natural within the formal framework presented in the subsequent
Section. However, it is less obvious on an intuitive level, especially
so since the reactance matrix is used much less frequently in the
scattering theory proper than its `cousin' the $T$-matrix (a very
detailed exposition of the subject can be found, however,
in~\cite{newton}). Nevertheless, it can be qualitatively understood
based precisely on the fact that reactance matrix is the Hermitian
analog of the $T$-matrix. Indeed, the latter is conventionally defined
by the equation $V |\chi_{\text{out}}\rangle = \mathcal{T}
|\chi_0\rangle$, where $|\chi_0\rangle$ is the
incident plane wave, and $|\chi_{\text{out}}\rangle$ is the scattered wave,
normalized to \emph{unit flux}. Similarly, $\mathcal{R}$ satisfies an
equation of identical structure, $V |\varphi_V\rangle =
\mathcal{R} |\varphi_0\rangle$, where
$|\varphi_V\rangle$ and $|\varphi_0\rangle$ are
now the conventionally normalized respectively, perturbed, and
unperturbed, states in a \emph{closed system}. Since we consider the
parametric correlations of the real eigenvalues of Hermitian
Hamiltonians in closed systems, reactance matrix is indeed quite a
natural object to appear.

The eigenvalues of $\mathcal{R}$ are expressed as tangents of the
corresponding eigenphases (scattering phase shifts) of $\mathcal{S}$. The
parameters $y_a$ introduced in Eq.~\eqref{eq:arctn} thus have an
intuitive physical interpretation as the corresponding phase
shifts. The expression of $x_a$ (or $y_a$) in terms of the reactance
matrix should be regarded as the analog of the microscopic definition
of $x_0$ given by Eq.~\eqref{eq:xexpl}. Moreover, the fact that the
'natural' parameters are the eigenvalues of the corresponding
reactance matrix allows for a more direct definition of the special
set of curves in the space of Hamiltonians which were discussed
above. These curves can now be represented as sequences of Hamiltonians
which map onto \emph{commuting} sequences of reactance matrices.


It is possible to complete the analogy with the case of
extended perturbations by establishing a counterpart of
Eq.~\eqref{eq:xdefi}. The distribution $P(u)$ of level velocities $u =
\partial \varepsilon_i/\partial |\vect{y}|$ can be extracted from
Eq.~\eqref{eq:sgene} with the help of the
identity~\cite{kravtsov_zirnbauer}:
\begin{equation}
  P(u) = \lim_{|\vect{y}| \to 0} |\vect{y}| R_{11}
  (\Omega/\bar{\Delta} = u |\vect{y}|, \vect{y}) \; .
\label{eq:kravz}
\end{equation}
Making use of this identity, one obtains the following result
\begin{equation}
  P(u) = \mathrm{Re} \int_0^\infty d\zeta\; e^{i \pi u_+ \zeta}\; e^{-
  \sum_{a}^r \ln (1 + i \zeta \kappa_a)} \; ,
\label{eq:leveu}
\end{equation} 
where $\kappa_a = y_a/|\vect{y}|$. Thus, the cumulants $C_m$ of the
level velocity are given by 
\begin{equation}
  C_m = \Frac{(m-1)!}{\pi^m} \sum_{a=1}^r \kappa_a^m \; .
\label{eq:cumul}
\end{equation}
By choosing $|\vect{y}|$ as the variable, we have already effectively
performed a rescaling analogous to~\eqref{eq:xdefi} (up to a factor of
$\pi$), with $C_2 \equiv C(0)$. Any set of additional $r-1$ cumulants
is sufficient to invert Eq.~\eqref{eq:cumul}, thus providing the
remaining (non-linear) variable transformations between
phenomenological constants $C_m$ and universal parameters
$y_a$. Equation~\eqref{eq:cumul} establishes the universality of
Eq.~\eqref{eq:sgene}.

Although we have kept the discussion above rather general, the most
interesting application from the practical point view is when $V$ is
diagonal in the coordinate representation, and $H_0$ and
$H_{\text{dis}}$ together describe weakly disordered metals. The
latter are characterized by the hierarchy of scales $L \gg \ell \gg
\lambda_F$. In this case it is convenient to distinguish two subtypes
of local perturbations $V$. In the first subtype, the spatial extent
$l_V$ of $V$ is smaller than $\ell$. For such perturbing potentials,
Eq.~\eqref{eq:sgene} can be more compactly rewritten as
\begin{equation}
  \sigma_{\text{loc}} (\lambda,\lambda_1;\vect{x}) = \tr \left(1 + i
  \mathcal{R} \lambda_1\right) \left(1 + i\mathcal{R}
  \lambda\right)^{-1}\; ,
\label{eq:sgen2}
\end{equation}
where $\mathcal{R}$ is the reactance matrix projected onto the Fermi
surface, and $\tr$ denotes the trace in the space of the scattering
channels. The full reactance matrix is defined by the equation
\begin{equation*}
  \mathfrak{R} = (1 - V \mathrm{Re}\langle G^{\smc{r}}\rangle)^{-1} V
  \; , 
\end{equation*}
where the retarded average Green function is defined as $\langle
G^{\smc{r}}\rangle \equiv \langle (\varepsilon_F - H_0 -
H_{\text{dis}} + i0)^{-1}\rangle$, so that $\mathfrak{R}$ is formally
distinct from the \emph{average reactance matrix} $\langle(1 - V
\mathrm{Re} G^{\smc{r}})^{-1}V \rangle$; nevertheless the difference
between $\mathfrak{R}$ and the average reactance matrix enters only in
the subleading order in $\lambda_F/\ell$, and can be
ignored. Transforming to the momentum representation it is convenient
to split off the angular variables in $\mathfrak{R}$ as $\mathfrak{R}
(\vect{p} , \vect{p}') = \mathfrak{R} (\xi,\hat{\vect{n}};
\xi',\hat{\vect{n}}')$, where $\xi = \vect{p}^2/2m -
\varepsilon_F$. The projected reactance matrix $\mathcal{R}$ is
defined as $\mathcal{R}_{\hat{\vect{n}} , \hat{\vect{n}}'} \equiv
\pi \bar{\nu} \mathfrak{R}(0,\hat{\mathbf{n}};0,\hat{\mathbf{n}}')$.

In the opposite limit $L \gg l_V\gg \ell$, Eq.~\eqref{eq:sgen2}
retains its structure provided $\mathcal{R}$ is replaced with
\begin{gather*}
  \mathcal{R}_P = P^{1/2}(\xi) \mathfrak{R} (\xi,\hat{\vect{n}} ;
  \xi',\hat{\vect{n}}') P^{1/2}(\xi')\; ,\\
\intertext{where}
  P(\xi) = \Frac{(1/2\pi\tau)}{\xi^2 + (1/2\tau)^2}\; ,
\end{gather*}
$\tau=\ell/v_F$ is the mean free time, and the trace operation $\tr$
is redefined to include the $\xi$ degrees of freedom.  In other words,
the projection onto the Fermi surface is smeared over a strip of width
$1/2\tau$ around it, reflecting the fact that the \emph{average} Green
function exponentially decays over distances larger than $\ell$ so
that a scatterer of size larger than $\ell$ is effectively split into
independent chunks of size $\ell$. Since the algebraic structure of
the parametric correlation functions is identical in both cases $l_V
<\ell$ and $l_V>\ell$, below we will often omit the subscript $P$.


In summary, the two ways of defining $\vect{y}$ either in terms of the
solutions of Eq.~\eqref{eq:cumul} or as the eigenphases of the
reactance matrix stand in direct correspondence with the definition of
$x_0$ either phenomenologically~\eqref{eq:xdefi} or
microscopically~\eqref{eq:xexpl}. While very convenient for
theoretical analysis, the definition in terms of the eigenphases is of
little practical utility since measuring $\mathcal{R}$ is hardly
feasible. On the other hand, cumulants of the level velocity
distribution are readily accessible in appropriately designed
experiments.

The universality of the parametric correlation functions is explicitly
demonstrated through the existence of their \emph{translationally
invariant parametrization in terms of the scattering phase shifts.}
In this context, the standard universal Wigner-Dyson statistics of
energy levels (\emph{spectral points}) could be viewed as the
\emph{boundary condition} to the universal statistics of
\emph{spectral lines} as functions of coordinates in the space of
Hamiltonians.

\subsection{Orthogonal ensemble}
\label{sec:ortho}
All of the preceding discussion can be applied verbatim to the case of
orthogonal symmetry provided Eqs.~\eqref{eq:sline}
and~\eqref{eq:sgen2} are replaced with
\begin{widetext}
\begin{equation}
  R_{11}^{(o)} (s , V) = 1 + \mathrm{Re} \int_{1}^{\infty} d\lambda_1
  d\lambda_2 \int_{-1}^{1} d \lambda\; J(\lambda,\lambda_1,\lambda_2)
  e^{i \pi s_+ (\lambda_1 \lambda_2 - \lambda) - \sigma^{(o)}
  (\lambda,\lambda_1,\lambda_2 ; x_0, \mathbf{x})} \; ,
\label{eq:ranko}
\end{equation}
\end{widetext}
where 
\begin{equation*}
  J(\lambda,\lambda_1,\lambda_2) = \Frac{(\lambda_1 \lambda_2 -
  \lambda)^2 (1 - \lambda^2)}{(\lambda_1^2 + \lambda_2^2 + \lambda^2 -
  2 \lambda \lambda_1 \lambda_2 - 1)^2}
\end{equation*}
and the global and local parts of $\sigma^{(o)}$ are given by,
respectively,
\begin{equation*}
  \sigma_{\text{gl}}^{(o)} (\lambda,\lambda_1,\lambda_2 ; x_0) =
  \Frac{\pi^2 x_0^2}{2} \left(1 + 2\lambda_1^2 \lambda_2^2 - \lambda^2
  -\lambda_1^2 -\lambda_2^2\right)
\end{equation*}
and
\begin{multline}
  \sigma_{\text{loc}}^{(o)} (\lambda,\lambda_1,\lambda_2 ; \mathbf{x})
  \\
  = \Frac{1}{2} \tr \ln \Frac{\openone + 2i \mathcal{R} \lambda_1
  \lambda_2 - \mathcal{R}^2(\lambda_1^2 + \lambda_2^2 - 1)}{(\openone
  + i\mathcal{R}\lambda)^2}\; .
\label{eq:sorth}
\end{multline}
As before, $\mathcal{R}$ has to be replaced with $\mathcal{R}_P$ with
the corresponding redefinition of $\tr$ if $l_V\gg \ell$.

A similar set of expressions can also be written in the symplectic
($\beta=4$) case.

\subsection{Level velocity distribution and Berry's conjecture}
\label{sec:berry}
Applying Eq.~\eqref{eq:kravz} to Eq.~\eqref{eq:ranko}, we find that
Eq.~\eqref{eq:leveu} is a special case of a more general expression:
\begin{equation}
  P (u) = \Frac{\beta}{2} \mathrm{Re} \int_0^\infty d\zeta\; e^{i \pi
  \beta u_+ \zeta/2}\; e^{- (\beta/2) \sum_{a}^r \ln (1 + i \zeta
  \kappa_a)} \; .
\label{eq:genvd}
\end{equation}
It is interesting to note that the distribution of level velocities
can be obtained using a completely different approach. As can be
easily seen from the expansion of an arbitrary eigenvalue
$\varepsilon_i$ of $H + XV$ to linear order in $X$,
\begin{gather*}
  \varepsilon_i (X) \simeq \varepsilon_i + X V_{ii} +O(X^2) \\ 
  V_{ii} = \int d\vect{r}\; |\psi_i (\vect{r})|^2 V(\vect{r}) \; ,
\end{gather*}
the level velocity coincides with the expectation value of $V$ in the
$i$-th eigenstate of $H$. Restricting our attention for simplicity to
the case $l_V \ll \ell$, we can employ Berry's conjecture about the
distribution of wave function values in chaotic
systems~\cite{berry,gornyi_mirlin}:
\begin{equation*}
  \mathcal{P} [\psi] = \Frac{1}{(\det \hat{f}^{-1})^{\beta /2}} e^{-
  (\beta/2) \int d\vect{r} d\vect{r}' \psi^\dag (\vect{r})
  [\hat{f}^{-1}]_{\vect{r} \vect{r}'} \psi (\vect{r}')} \; ,
\end{equation*}
where in $d$ spatial dimensions the matrix elements of the integral
operator $\hat{f}$ are given by the Friedel function~\eqref{eq:ffrie},
which in the limit $l_V \ll \ell$ can be approximated as $f_d
(\vect{r} - \vect{r}') \equiv \int d\vect{n}\; e^{ip_F \vect{n}\cdot
(\vect{r} - \vect{r}')} = J_{d/2-1}(p_Fr)$. In this expression
the integration is performed over all directions of the unit vector
$\vect{n}$ and $r=|\vect{r} - \vect{r}'|$.  The fields $\psi$ are
complex in the unitary case, and real for $\beta =1$. While originally
\emph{conjectured} for chaotic systems~\cite{berry}, this expression
was recently \emph{proved}~\cite{gornyi_mirlin} to hold locally in
diffusive systems of unitary symmetry.  We thus immediately find
\begin{multline}
  P (u) = \bar{\Delta} \sum_{i} \langle \delta (u -
  V_{ii}/\bar{\Delta}) \delta (\varepsilon - \varepsilon_i)\rangle \\  
  = \int_{-\infty}^{+ \infty} \Frac{d \zeta}{2\pi}\int D\psi_i
  \mathcal{P} [\psi_i]\; e^{i u \zeta - i \zeta \int d\vect{r}\; |\psi_i
  (\vect{r})|^2 V (\vect{r})/\bar{\Delta}} \\
  = \int_{-\infty}^{\infty} \Frac{d \zeta}{2\pi}\; e^{i\zeta u}
  \Frac{1}{[\det (\openone + 2i \zeta \hat{f} V/\beta
  \bar{\Delta})]^{\beta/2}} \;  .
\label{eq:lvber}
\end{multline} 
The last expression can be shown to coincide with
Eq.~\eqref{eq:genvd}, thus indirectly confirming the validity of
Berry's conjecture in diffusive systems of orthogonal
symmetry. Indeed, $y_a/|\vect{y}|$ coincide with the eigenvalues of
the Fermi surface projection of $d\mathcal{R}/dX|_{X\to 0}\equiv
V$. The corresponding projected matrix elements are $\int d\vect{r}\;
e^{ip_F (\vect{n} - \vect{n}') \cdot \vect{r}} V(\vect{r})$. Using the
integral decomposition of $f_d(\vect{r} -\vect{r}')$ and the cyclic
invariance of the determinant, the equivalence of
Eqs.~\eqref{eq:genvd} and~\eqref{eq:lvber} is immediately
established. This analysis can also be in an obvious way extended to
the case $L\gg l_V \gg \ell$.

\begin{figure}
\begin{center}
\includegraphics[width=1\linewidth,angle=0]{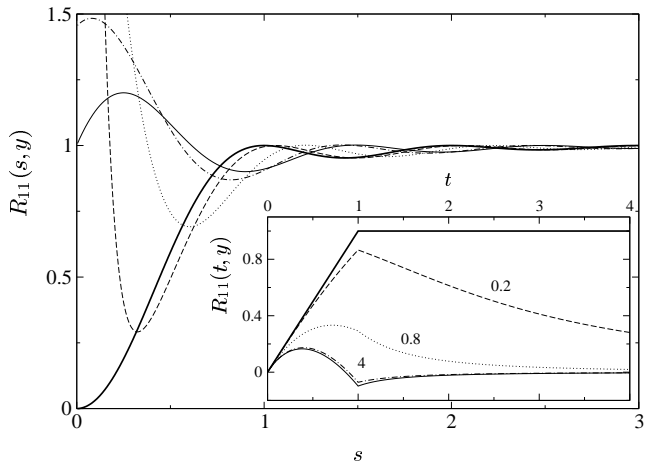}
\end{center}
\caption{\small DoS-DoS correlation function for the rank $r=1$
        problem, $R_{11} (s,y)$, for different strengths $\tan y$ of
        the perturbing potential: $\tan y=0.2$ (dashed line), $\tan
        y=0.8$ (dotted line), $\tan y=4$ (dot-dashed line), $\tan y
        \to \infty$ (solid line), while the bold line corresponds to
        the GUE Wigner-Dyson result, $y=0$. The corresponding Fourier
        transforms are shown in the inset.}
\label{fig:matve}
\end{figure}
%

\subsection{Examples}
\label{sec:examp}
Before turning to derivation of the above results, it is instructive
to analyze several special cases. For example, if $V$ models a
collection of several impurities, it can be approximated as 
\begin{equation}
  V(\vect{r}) = \Frac{1}{\bar{\nu}} \sum_{k=1}^r v_k \delta
  (\vect{r}-\vect{r}_k)\; .
\label{eq:locav}
\end{equation}
Crudely speaking, a bistable impurity would have $r=2$ and $v_1=-v_2$,
while a local defect created, e.g., by irradiation would correspond to
$r=1$. 

Throughout this subsection we largely restrict the discussion to the
unitary ensemble. The corresponding results for $\beta=1$ can be
straightforwardly obtained in terms of the corresponding $\lambda$
integrals starting from Eq.~\eqref{eq:sorth}.

Despite the fact that $\mathcal{R}$ (or $\mathcal{R}_P$) is a
continuous integral operator, the structure of Eq.~\eqref{eq:locav}
ensures that it possesses exactly $r$ non-zero eigenvalues. Using the
cyclic invariance of the trace, the latter are easily seen to coincide
with the eigenvalues of the asymmetric matrix $M_{kk''} =
\sum_{k'=1}^r \mathcal{R} (\vect{r}_k , \vect{r}_{k'}) f_d
(\vect{r}_{k'} - \vect{r}_{k''})$, where, for arbitrary $l_V \ll L$,
the Friedel function $f_d(\vect{r})$ is given by
Eq.~\eqref{eq:ffrie}.

\begin{figure} 
\begin{center} 
\includegraphics[width=1\linewidth,angle=0]{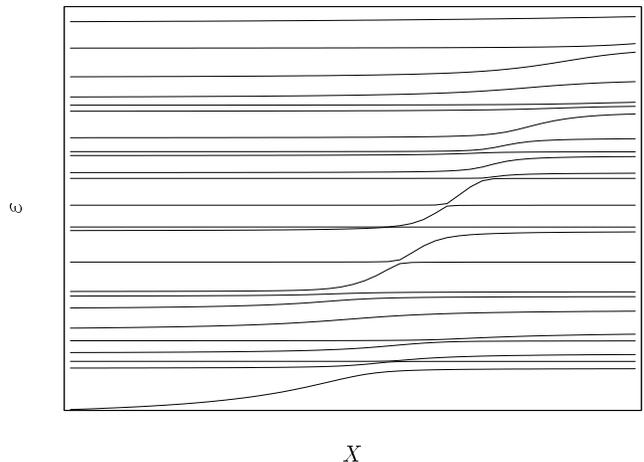}
\end{center}
\caption{\small Dependence of the energy levels on the overall
         strength $X$ of an external \emph{local} (i.e.,
         localized on a single site) perturbation $V$ for a
         time-reversal invariant tight-binding $20 \times 20$ lattice
         model (arbitrary units). Level anti-crossing and the property
         $\varepsilon_i < \varepsilon_i (X) <
         \varepsilon_{i+1}$~\cite{aleiner_matveev} are clearly shown.}
\label{fig:levr1}
\end{figure}

In the simplest $r=1$ case, $M_{kk'}$ is a number, $\tan y= 
v/(1-\alpha v)$, where for brevity we denote $\alpha = \mathrm{Re}
\langle G^{\smc{r}} (\vect{r} = 0)\rangle/\bar{\nu}$, and the
correlation function in the unitary ensemble can be explicitly written
in the form (assuming $y>0$)
\begin{multline}
  R_{11} (s , y) - 1 = - \left\{\Frac{\partial}{\partial s}
  \left[e^{-\pi s\cot y} k(s)\right]\right\}\\ 
  \left[\theta(s) - \int_{-\infty}^{s} ds' e^{\pi s'\cot y}
  k(s')\right]\; ,
\label{eq:rkone}
\end{multline}
where the function $k(s)$ has been defined in~\eqref{eq:kfunc},
therefore reproducing the result obtained
previously~\cite{aleiner_matveev} using RMT methods.  The function
$R_{11}$ for $r=1$ is plotted in Fig.~\ref{fig:matve} for several
different values of $y$ together with its Fourier transform. One can
clearly see the gradual broadening of the central peak inherited from
the $\delta$-function term in Eq.~\eqref{eq:rmtbe}. At small values of
$y$ the dominant contribution to the broadened peak comes from the
correlations between a level and its parametric `descendant'. It is
also worth noting that the parametric correlation function inherits
from its non-parametric limit the sharp oscillatory behavior which is
reflected in the singularity (cusp) in its Fourier transform
\begin{multline*}
  R_{11} (t , y) = \int_{-\infty}^{\infty} ds\; e^{-2\pi i s t} R_{11}
  (s , y)\\ 
  = \delta(t) + \min(t , 1) + \Frac{t}{2} \ln \left[\Frac{1 + \tan^2 y
  g^2 (t)}{1 + \tan^2 y (1 + 2t)^2}\right] \; ,
\end{multline*}
where $g(t) = 2t - 2\min(t,1) +1$. Although not directly obvious from
Eq.~\eqref{eq:rkone}, the perturbed levels in the $r=1$ case
possess the property (for $y>0$) $\varepsilon_i < \varepsilon_i(y) <
\varepsilon_{i+1}$ \cite{aleiner_matveev}. This feature is clearly
illustrated in Fig.~\ref{fig:levr1}.

\begin{figure}
\begin{center}
\includegraphics[width=1\linewidth,angle=0]{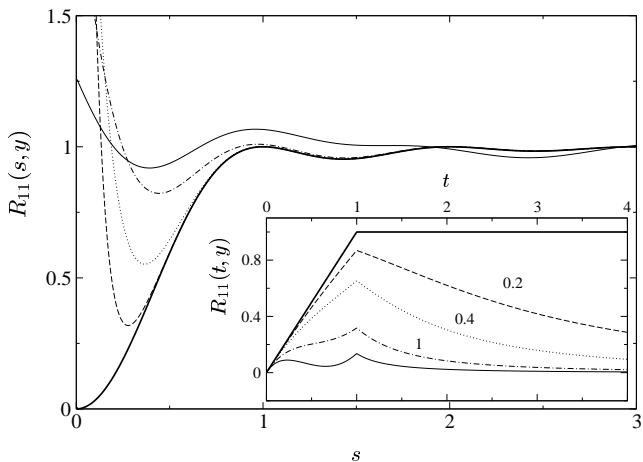}
\end{center}
\caption{\small DoS-DoS correlation function for the rank $r=2$
        problem, $R_{11} (s,y)$, for different values of $\tan y =
        \tan y_1 = -\tan y_2$. The bold line corresponds to the GUE
        Wigner-Dyson result, $\tan y=0$, while $\tan y=0.2$ (dashed
        line), $\tan y=0.4$ (dotted line), $\tan y=1$ (dot-dashed
        line), $\tan y \to \infty$ (solid line). The corresponding
        Fourier transforms are shown inset.}
\label{fig:onemo}
\end{figure}

In the $r=2$ case we restrict our attention to modeling a bistable
impurity, thus setting $v_1 = -v_2 = v >0$, and $r_{12} \ll
\ell$. Denoting as before $\alpha = \mathrm{Re} \langle G^{\smc{r}}
(\vect{r} = 0)\rangle/\bar{\nu}$, and also $\gamma = \mathrm{Re}
\langle G^{\smc{r}} (r_{21})\rangle/\bar{\nu}$,
$r_{kk'}=|\vect{r}_{k}-\vect{r}_{k'}|$, and $f_{kk'} = f_d(r_{kk'})$,
we find
\begin{equation*}
  \tan y_{1,2} = \Frac{v^2 (\alpha - \gamma f_{12}) \pm v \sqrt{1 -
      f_{12}^2 + v^2 (\alpha f_{12} - \gamma)^2}}{1 - v^2 (\gamma^2 -
      \alpha^2)} \; .
\end{equation*}
In the limit $r_{12} \to 0$ both eigenvalues, as expected, vanish. We
omit somewhat lengthy explicit expressions for $R_{11}$ and its
Fourier transform, presenting instead in Fig.~\ref{fig:onemo} the
corresponding graphs at several different values of $y$. The limit
$r_{12}\to 0$ can also be used to extract the distribution of level
velocities due to a moving impurity. Expanding $1 - f_{12} \approx
(1/2 d)p_F^2 r^2$ we find (setting for simplicity $\alpha = \gamma =
0$) that the distribution of $u = \partial\varepsilon_i/\partial (p_F
r)$ is Poissonian:
\begin{equation}
  P(u) = \Frac{\sqrt{d}}{2v} e^{-|u|\sqrt{d}/v} \; .
\label{eq:defve}
\end{equation}

This result can be generalized to study the response of the energy
levels to a shift in the position of an extended defect. An
experimentally relevant application is the lateral motion of an STM
tip over a disordered two-dimensional electron gas. Another example is
a metallic scatterer inside a microwave `billiard' such as those
studied in Ref.~\cite{sridhar}.

\begin{figure}
\begin{center}
\includegraphics[width=1\linewidth,angle=0]{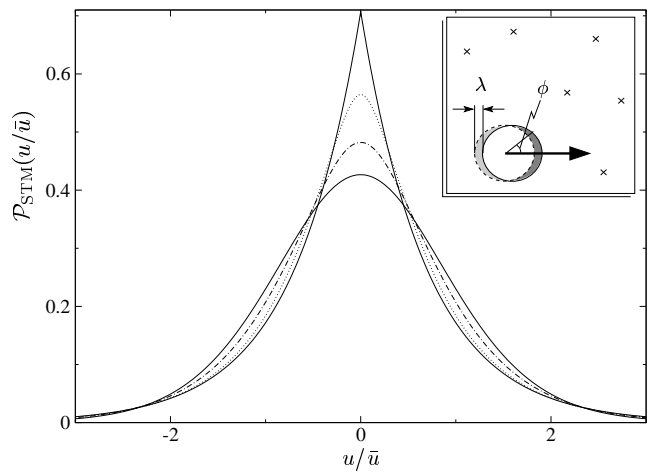}
\end{center}
\caption{\small Graph of $P_{\smc{stm}}(u/\bar{u})$ in the unitary
        ensemble obtained for $p_F\rho=1.5$ and $4.0$ together with
        the limiting cases of Poisson ($p_F\rho \rightarrow 0$) and
        Gaussian ($p_F\rho\rightarrow\infty$) distributions. The inset
        shows the perturbation $V(\vect{r})$ arising from the
        displacement of the disk-shaped potential.}
\label{fig:veuni}
\end{figure}

\begin{figure}
\begin{center}
\includegraphics[width=1\linewidth,angle=0]{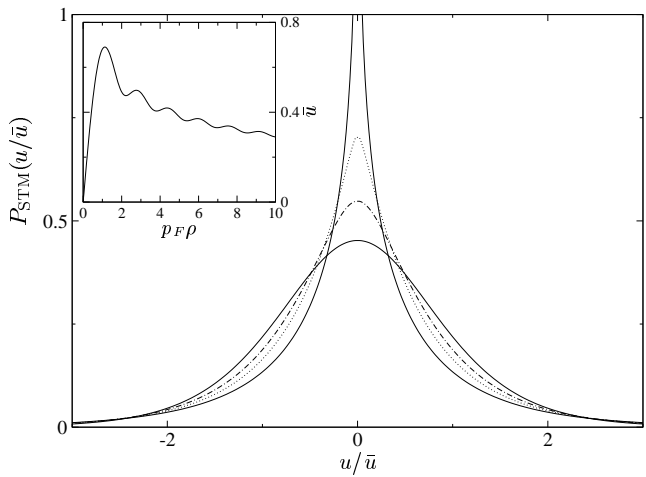}
\end{center}
\caption{\small Graph of $P_{\smc{stm}}(u/\bar{u})$ in the orthogonal
        ensemble obtained for $p_F\rho=1.8$ and $p_F\rho=5$ together
        with the limiting cases of $p_F\rho \rightarrow 0$ and
        $p_F\rho\rightarrow\infty$. The inset shows $\bar{u}=\langle
        u^2\rangle^{1/2}$ as a function of $p_F\rho$.}
\label{fig:movfi}
\end{figure}

Approximating the potential $U$ of an STM tip as a flat disk of radius
$\rho$, $U(\vect{r}) = U_0 \theta (\rho^2 - \vect{r}^2)$, and denoting
the displacement of the center of the disk as $\lambda$, the
difference between the potentials produced at two adjacent positions
of the disk is given, to the first order in $\lambda$, by $V(\vect{r})
\equiv V(r , \phi) = 2 U_0 \lambda \rho \delta (\vect{r}^2 - \rho^2)
\cos \phi$, where the direction of the displacement corresponds to
$\phi=0$.  Therefore, neglecting the $O(U_0)$ corrections to
$f_d(\vect{r}) = J_0 (p_F r)$, one finds that the operator $\hat{f}
V/\bar{\Delta}$ appearing in the expression for the velocity
distribution~\eqref{eq:lvber} reduces to a continuous integral
operator defined on a circle $0 \le \phi < 2\pi$:
\begin{equation*}
  \kappa(\phi , \phi') = \Frac{1}{2\pi} J_0(2 p_F \rho
  \sin((\phi-\phi')/2))\cos \phi' \; .
\end{equation*}
Since the corresponding eigenvalues $\kappa_a$ are symmetric with
respect to zero, the distribution of level velocities can be written
as
\begin{equation*}
  P_{\smc{stm}} (u/\bar{u}) = \int_{- \infty}^{\infty}
  \Frac{d\zeta}{2\pi} e^{i \zeta (u/\bar{u}) - (\beta /2)
  \sum_{\kappa_a>0} \ln [1 + (2 \kappa_a \zeta/\beta \bar{u})^2]} \; ,
\end{equation*}
where the average velocity $\bar{u}$ is given by
\begin{multline*}
  \bar{u}^2 \equiv \langle u^2 \rangle = \Frac{2}{\beta} \sum_a
  \kappa_a^2 \\
  = \Frac{1}{\beta\pi} \int_0^\pi d\theta J_0^2 (2p_F \rho \sin
  \theta) \cos (2 \theta )\; .
\end{multline*}
In the unitary case, $\beta = 2$, the distribution function
(Fig.~\ref{fig:veuni}) shows a crossover from Poisson to Gaussian
behavior as $p_F\rho$ increases, while in the orthogonal case
(Fig.~\ref{fig:movfi}) the crossover is between the limiting behavior
described by the modified Bessel function $(1/\pi) K_0(|u|)$
\cite{barth} and the Gaussian limit. These crossovers explicitly
illustrate the distinction between local and global perturbations, and
show that the global regime is achieved when the central limit theorem
comes into force due to a large number of distinct eigenvalues
$\kappa_a$.  Such a crossover has been observed in experiments on
microwave resonators~\cite{barth} where the appropriate ensemble is
orthogonal.

A new feature appearing in this example is that a local potential is
described by a formally infinite number of the eigenvalues $\kappa_a$
of an integral operator. However, the finite extent of $V$ guarantees
that all but a finite number of these eigenvalues are vanishingly
small, so that, depending on the required degree of accuracy, there
can always be defined an appropriate \emph{finite} value of $r$.

\section{Field Theory of Parametric Correlations}
\label{sec:ftheo}
We turn now to the derivation of the results presented in the
preceding Section.  To explore the influence of a local potential
perturbation on the ensemble average properties of the weakly
disordered system, we will employ a conventional approach based on the
supersymmetric field theoretic formulation. Since this approach has
been reviewed extensively in the literature~\cite{efetov}, we will
keep our discussion here concise, paying particular attention only to
the idiosyncrasies of the present scheme.


Focusing on the weakly disordered system, our starting point is the
set of single-particle Hamiltonians
\begin{equation*}
  H_i = \hat{\xi}_{\hat{\vect{p}} - e\vect{A}} + U(\vect{r}) +
  V_i (\vect{r}) \; ,
\end{equation*}
where $\hat{\xi}_{\hat{\vect{p}} - e\vect{A}} = (\hat{\vect{p}} -
e\vect{A})^2 /2m - \varepsilon_F$ is the kinetic energy operator a
free-particle system subject to an external vector potential
$\vect{A}$.  The impurity potential $U(\vect{r})$ is drawn at random
from a Gaussian $\delta$-correlated distribution with zero mean, and
variance
\begin{equation}
  \langle U(\vect{r}) U(\vect{r}') \rangle = \Frac{1}{2 \pi \bar{\nu}
  \tau} \delta (\vect{r} - \vect{r}') \; ,
\label{eq:varia}
\end{equation}
where $\tau$ is the associated elastic mean free time, $\tau = \ell/
v_F$. In the following, we will limit our considerations to the
diffusive regime, where the sample size $L$ is greatly in excess of
the mean free path $\ell$ and where the wavefunction is extended over
the volume of the system. Moreover the energy scales are arranged in
the hierarchy
\begin{displaymath}
  \varepsilon_F \gg \Frac{1}{\tau} \gg E_c \gg \bar{\Delta} \; ,
\end{displaymath}
with $E_c = D/L^2$ and $D = v_F^2 \tau/d$. In addition to the random
potential $U(\vect{r})$, the diffusive system is subject to a further
(potentially short-ranged) arbitrary external parametric perturbation
which can take $n$ values $V_i(\vect{r})$.

Our goal initially is to construct a formalism which allows, at least
in principle (see below), to compute multi-point correlation functions
of the DoS,
\begin{equation}
  R(\vectgr\varepsilon ; \vect{V}) = \langle \prod_{i=1}^p \nu
  (\varepsilon_i ; V_i) \rangle\; ,
\label{eq:multi}
\end{equation}
where $\vectgr\varepsilon$ and $\vect{V}$ denote $p$-dimensional
vectors with components $\varepsilon_i$ and $V_i$, respectively, and
$\nu(\varepsilon_i;V_i)$ is the value of the density of states at
energy $\varepsilon_i$ of the Hamiltonian $\hat{\xi}_{\hat{\vect{p}}}
+ U + V_i$. For generality, let us assume that $p\ge n$, so that
some of $V_i$ in ~\eqref{eq:multi} are equal. The $p$ energies in
$\vectgr\varepsilon$ (which \emph{are} generically different) are
correspondingly split into $n$ groups of size $p_j$, each group
matching a given value $V_j$. Moreover, from now on, we assume
that the Fermi energy $\varepsilon_F$, which is included in the free
Hamiltonian $\hat{\xi}_{\hat{\vect{p}}}$, is subtracted from the
energies $\varepsilon_i$.

According to the standard methods
\cite{efetov,altshuler_simons},
the generating functional required to construct a field-theoretic
representation of $R(\vectgr\varepsilon;\vect{V})$ in the case of unbroken
time-reversal symmetry is written as a functional integral over an
$8p$-dimensional supermultiplet of complex fields:
\begin{widetext}
\begin{equation}
  \mathcal{Z}[j] = \int D(\psi^\dag,\psi) \exp \left\{i \int d\vect{r}
  \left\{\bar{\Psi} \left[\widehat{\varepsilon}^{\;c} -
  \hat{\xi}_{\hat{\vect{p}}} - U(\vect{r})- \widehat{V}
  (\vect{r})\right] \Psi + \Psi^\dag j + j^\dag \Psi\right\}\right\}
  \; ,
\label{eq:sourc}
\end{equation}
\end{widetext}
where, by choosing the fields $\psi$ to consist of $2p$ copies of both
boson ($\smc{b}$) and fermion ($\smc{f}$) elements, the normalization
of the generating function, $\mathcal{Z}[0] = 1$ is enforced. The
factor $2$ in $2p$ is explained by the need to generate both retarded
($\smc{r}$) and advanced ($\smc{a}$) Green functions, so that
$\varepsilon_i^{c} = (\varepsilon_i \sigma_0^{\smc{ra}} +
\sigma_3^{\smc{ra}} i0) \otimes \sigma_0^{\smc{bf}} \otimes
\sigma_0^{\smc{tr}}$. The further doubling of the number of components
of the superfields is dictated by the need to properly take into
account the soft modes associated with the time-reversal invariance of
the Hamiltonian (hence the notation $\smc{tr}$ for the corresponding
subspace):
\begin{align*}
  \Psi &= \Frac{1}{\sqrt{2}} \begin{pmatrix} 
  \psi \\ 
  \psi^* \end{pmatrix}_{\smc{tr}} & & \begin{split}
  \Psi^\dag &= \left(C \Psi\right)^{\mathsf{T}} \\ 
  \Psi &= C (\Psi^\dag)^{\mathsf{T}} \; ,
  \end{split}
\end{align*}
where $C = \sigma_1^{\smc{tr}} \otimes E_{11}^{\smc{bf}} + i
\sigma_2^{\smc{tr}} E_{22}^{\smc{bf}}$, and the $E$-matrices are the
projectors onto the corresponding parts of the superspace.  To stay
close to the conventional notation adopted in the literature, in the
following we will denote $\sigma_3^{\smc{ra}} \equiv \Lambda$. The
conjugate superfields are defined by $\bar{\psi} = \psi^\dag L$, where
$L = \Lambda \otimes E_{11}^{\smc{bf}} + \openone \otimes
E_{22}^{\smc{bf}}$ (further details on notation can be found
in~\cite{efetov}). Finally, $\widehat{\varepsilon}^{\;c}=
\diag(\varepsilon_1^{c} , \dots , \varepsilon_p^{c})$, $\widehat{V} =
\diag (V_1 , \dots , V_p)$, and the absence of an explicit operator in
any subspace always implies the corresponding identity operator.

The only exception to this structure is the case $p=2$, where it is
sufficient to represent each of the two DoS factors by either a
retarded or an advanced Green function, with the opposite choice for
the remaining factor, thus reducing the number of required components
of the superfields from $16$ to $8$.

An ensemble average of the generating functional $\mathcal{Z}[j]$ over
realizations of the random impurity potential $U(\vect{r})$ induces a
quartic interaction of the fields $\Psi$, which can be decoupled by
means of a Hubbard-Stratonovich transformation with the introduction
of $8p \times 8p$-component supermatrix fields $\mathcal{Q}
(\vect{r})$. In the absence of symmetry breaking sources, the
resulting action is invariant under pseudounitary transformations,
$\mathcal{Q} \mapsto T \mathcal{Q} T^{-1}$, where $T$ satisfies
$T^\dag L T = L$, and thus belongs to the pseudounitary supergroup
$\mathrm{U} (2p, 2p|4p)$. Moreover, $\mathcal{Q}$ satisfies the
time-reversal symmetry constraint $\mathcal{Q} = L C^{\mathsf{T}}
\mathcal{Q}^{\mathsf{T}} C L$. 

After performing the integration over the superfields $\Psi$, one
obtains the ensemble-averaged generating functional 
\begin{equation*}
  \langle\mathcal{Z} [0]\rangle = \int D\mathcal{Q}\;
  e^{-S[\mathcal{Q}]}\; ,
\end{equation*}
where the action $S[\mathcal{Q}]$ is
\begin{gather}
\label{eq:primo}
  S[\mathcal{Q}] = -\Frac{\pi \bar{\nu}}{8 \tau} \int d\vect{r} \;
  \str \mathcal{Q}^2 + \frac{1}{2} \int d\vect{r} \; \str \langle
  \vect{r}| \ln \hat{\mathcal{G}}^{-1} |\vect{r}\rangle \; ,
\intertext{and}
  \mathcal{G}^{-1}[\mathcal{Q};\widehat{V}] =
  \widehat{\varepsilon}^{\;c} - \hat{\xi}_{\hat{\vect{p}}} - \widehat{V}
  (\vect{r}) + \Frac{i}{2\tau} \mathcal{Q}(\vect{r})
\nonumber
\end{gather}
denotes the supermatrix Green function. The description of the
structure of supermatrices and the definition of the supertrace
(here, $\str = \tr_{\smc{b}} - \tr_{\smc{f}}$) operation can be found 
in~\cite{efetov}. 

The action~\eqref{eq:primo} possesses an almost degenerate
saddle-point manifold. Varying the action with respect to
$\mathcal{Q}$, one obtains the saddle-point equation
\begin{equation}
  \mathcal{Q} (\vect{r}) = \Frac{i}{\pi \bar{\nu}} \langle \vect{r}|
  \mathcal{G} [\mathcal{Q} ; \widehat{V}] |\vect{r} \rangle \;
  .
\label{eq:saddl}
\end{equation}
This equation can be interpreted as the self-consistent Born
approximation for the self-energy of the supermatrix Green
function. The ambiguity involved in choosing among the different
disconnected solutions of this equation is resolved by taking into
account the analytical properties of the average Green function:
$\mathcal{Q} = \Lambda$. In the limit $\widehat{\varepsilon} \to 0$ and
$\widehat{V} \to 0$, the saddle-point solution expands to fill the
degenerate manifold generated by transformations $\mathcal{Q} =
T\Lambda T^{-1}$, where $T \in \mathrm{U} (2p , 2p|4p)$.

\subsection{The Non-Linear $\sigma$ Model}
\label{sec:massi}
In the standard scheme~\cite{efetov} leading to the NL$\sigma$M
structure of the field theory, the fluctuations of $\mathcal{Q}$ in
the direction perpendicular to the saddle-point manifold are massive
due to the inequality $p_F \ell \gg 1$. Moreover, in the leading order
in $1/p_F\ell$, these fluctuations are quadratic and independent of
the transverse fluctuations. As a result, integration over the massive
modes does not lead to any modifications of the saddle point action
apart from an overall constant multiplying $\mathcal{Z}$;
supersymmetry further ensures that the constant is equal to $1$. In
the case of global parametric perturbations this scheme is preserved
\cite{altshuler_simons,simons_altshuler} due to
the fact that $V(\vect{r})$ is locally small, as discussed
above. However, it is not \emph{a priori} obvious that the same is
true in the case of local perturbations, since the latter are locally
large. Therefore, care has to be exercised in the derivation of
the parametric version of NL$\sigma$M to demonstrate that (i) local
perturbations do not destroy locally the saddle-point structure
(i.e. that the distinction between the massless transverse and massive
longitudinal fluctuations is preserved), and (ii) that any possible
cross-couplings between massive and soft modes mediated by $V$ do not
modify the saddle-point action in the leading order in $1/p_F\ell$.

To undertake this program, we begin by separating the fluctuations
around the saddle-point manifold into transverse modes $Q (\vect{r})$,
which are nearly massless and slowly varying on the scale of the mean
free path $\ell$, and massive modes $\delta Q_m (\vect{r})$, in such a
way that $\mathcal{Q} (\vect{r}) = Q (\vect{r}) + \delta Q_m
(\vect{r})$.  The latter include the longitudinal fluctuations $\delta
Q_l$ as well as \emph{fast} transverse fluctuations $\delta Q_f$. The
need to account for fast transverse fluctuations arises from the fact
that local perturbations vary rapidly on the scale of the wavelength
$\lambda_F$, so that their coupling to fast transverse modes cannot
\emph{a priori} be ignored. In principle, a na\"{\i}ve inclusion of
fast transverse fluctuations can lead to overcounting, as,
\emph{e. g.} a $2p_F$ Diffuson mode is a Cooperon, and vice
versa. Nevertheless, below we will demonstrate that massive modes do
not in fact generate any corrections to the slow mode action in the
leading order in $1/p_F\ell$, and a detailed calculation of the
subleading terms is not needed.

By definition, the longitudinal modes are orthogonal to the
saddle-point manifold, therefore satisfying $[\delta Q_l (\vect{r}), Q
(\vect{r})] = 0$. In contrast, the slow transverse modes can be
parametrized as $Q (\vect{r}) = T (\vect{r}) \Lambda T^{-1}
(\vect{r})$. The corresponding free supermatrix Green function,
$\mathcal{G}_0 [Q]= \mathcal{G} [\widehat{\varepsilon}^{\;c}=0 ,
\delta Q_m = 0; \widehat{V}=0]$, obeys the relation:
\begin{multline*}
  \mathcal{G}_0 (\vect{r},\vect{r}') = \mathrm{Re} \langle
  G^{\smc{r}} (\vect{r} , \vect{r}')\rangle \\
  - i \pi \bar{\nu} f_d (\vect{r} - \vect{r}') Q \left( \Frac{\vect{r}
  + \vect{r}'}{2}\right) \; ,
\end{multline*}
where the  retarded Green function is defined as $G^{\smc{r}} =
(-\hat{\xi}_{\hat{\vect{p}}} - U + i0)^{-1}$ and where $f_d (\vect{r} -
\vect{r}')$ denotes the Friedel function~\eqref{eq:ffrie}.

Separating the $V$-dependent parts of the action, we represent it as
\begin{equation*}
  S [\mathcal{Q}] = S[Q] + \delta S[Q,\delta Q_m] \; ,
\end{equation*}
where
\begin{multline}
  S[Q] = - \Frac{\pi \bar{\nu}}{8 \tau} \int d\vect{r}\; \str Q^2 +
  \Frac{1}{2} \int d\vect{r}\; \str \langle \vect{r}| \ln
  \mathcal{G}^{-1} [Q] |\vect{r} \rangle\\ 
  + \frac{1}{2} \int d\vect{r}\; \str \langle \vect{r}|
  \ln\left(\openone - \mathcal{G} [Q] \widehat{V}\right)|
  \vect{r} \rangle \; ,
\label{eq:sqact}
\end{multline}
$\mathcal{G}[Q] \equiv \mathcal{G} [Q ; \widehat{V}=0]$,
and the expansion
\begin{multline*}
  \mathcal{G} [\mathcal{Q}] = \mathcal{G}[Q] - \Frac{i}{2\tau}
  \mathcal{G} [Q] \delta Q_m \mathcal{G}[Q]\\
  -\Frac{1}{(2\tau)^2} \mathcal{G}[Q] \delta Q_m \mathcal{G}[Q] \delta
  Q_m \mathcal{G}[Q] 
\end{multline*}
determines the form of $\delta S[Q,\delta Q_m]$. In the limit
$\widehat{\varepsilon} \tau \ll 1$ the cross-coupling between
$\widehat{\varepsilon}^{\;c}$ and $\widehat{V}$ can be neglected, so that
$\mathcal{G}[Q]$ in the last term in (\ref{eq:sqact}) can be replaced
with $\mathcal{G}_0 [Q]$.

Employing the condition $\varepsilon_F \tau \gg 1$, a gradient
expansion of the first two terms in~\eqref{eq:sqact} leads to the
conventional non-linear $\sigma$-model action
\begin{equation*}
  S_0[Q] = -\Frac{\pi\bar{\nu}}{8} \int d\vect{r}\; \str \left[D
  (\nabla Q)^2 + 4i \widehat{\varepsilon} Q\right] \; .
\end{equation*}
Using the identity $\str\ln (\openone - \mathrm{Re}\left\langle
G^{\smc{r}}\right\rangle \widehat{V})=0$, the last term
in~\eqref{eq:sqact} can be rewritten as
\begin{equation}
  S_V [Q] = \frac{1}{2}\str \ln \left(\openone + i Q
  \widehat{\mathcal{R}}\right) \; ,
\label{eq:svfir}
\end{equation} 
where the operation $\str$ is assumed to include the trace over the
scattering channels degrees of freedom. Similarly, $\delta S[Q,\delta
  Q_m]$ takes the form
\begin{multline}
  \delta S[Q,\delta Q_m] = - \Frac{\pi\bar{\nu}}{8\tau} \str \delta
   Q_l^2 + \delta S_0[Q,\delta Q_f]\\
  +\Frac{1}{2} \str \ln \left\{\openone + \Frac{i}{2\pi \bar{\nu} \tau}
  \mathcal{G}_0 [Q] \delta Q_m \mathcal{G}_0[Q] \widehat{\mathcal{T}}
  \right.\\
  \left. -\Frac{1}{\pi\bar{\nu} (2\tau)^2} \mathcal{G}_0[Q]
   \delta Q_m \mathcal{G}_0[Q]\delta Q_m \mathcal{G}_0[Q]
   \widehat{\mathcal{T}}\right\}\; , 
\label{eq:delta}
\end{multline}
where
\begin{equation}
  \widehat{\mathcal{T}} = \widehat{\mathcal{R}} \left(\openone + iQ
  \widehat{\mathcal{R}}\right)^{-1}\; ,
\label{eq:gsmal}
\end{equation}
while $\delta S_0[Q,\delta Q_f]$ is generated from the high-order
terms in the gradient expansion, and can be approximated as $\str
\delta Q_f^2$ with a coefficient of the order of
$\bar{\nu}/\tau$. Both $Q_m$ and $\mathcal{G}_0[Q]$ in the above
expressions are taken in the momentum representation, with the
corresponding adjustment in the definition of $\str$. The matrix
$\widehat{\mathcal{T}}$ can be viewed as a supersymmetric
generalization of the $T$-matrix $\mathcal{T} \equiv (\mathcal{S} -
\openone)/2 i$, where $\mathcal{S}$ is the scattering matrix.

The crucial property of Eq.~\eqref{eq:delta} is that even if the
eigenvalues of $\widehat{\mathcal{R}}$ grow indefinitely,
$\widehat{\mathcal{T}}$ stays finite. Moreover, the Hermiticity of
$\mathcal{R}$ and $Q$ ensures that the denominator in~\eqref{eq:gsmal}
does not generate any singularities. Expanding the logarithm up to
quadratic order in $\delta Q_m$, and averaging
the action defined by the first two terms in the
r.h.s. of~\eqref{eq:delta} over $\delta Q_m$, we find that the corresponding
contributions to the slow mode action are small as
$1/p_F\ell$. Consequently, 
\begin{equation}
S[Q]=S_0[Q] + S_V[Q]
\end{equation}
represents the total slow
mode action in the leading order in $1/p_F\ell$.

This completes the formal construction of the non-linear
$\sigma$-model action. In the absence of an external perturbation $V$,
the functional integral is dominated by the coordinate-independent
zero mode $Q_0$ provided $\varepsilon_i \ll E_c$ for all $i$. In this
limit one recovers the familiar zero-dimensional non-linear
$\sigma$-model action~\cite{efetov},
\begin{equation}
  S_0[Q_0] = -\Frac{i\pi}{2\bar{\Delta}} \str[\widehat{\varepsilon}
  Q_0] \; , 
\label{eq:sigm0}
\end{equation}
which is well-known to reproduce the standard Wigner-Dyson correlation
functions~\cite{efetov}. The corresponding zero-mode contribution to
the action describing parametric correlations is given by
\begin{equation}
  S_V [Q_0] = - \Frac{1}{2} \str\ln\left(\openone +
  i\widehat{\mathcal{R}} Q_0\right) \; .
\label{eq:sinte2}
\end{equation}
As will be shown in the next subsection, Eq.~\eqref{eq:sinte2}
represents the dominant contribution to the action when $\widehat{V}$
is a local perturbation. In the opposite case, the $1/g$ corrections
to the action can compete with the zero-mode contribution, and their
relative importance depends on the spectral composition of
$\widehat{V}$. 

\subsection{Local versus Global Perturbations}
\label{sec:lovgl}
While the zero-dimensional non-linear $\sigma$-model~\eqref{eq:sigm0}
combined with the interaction action~\eqref{eq:sinte2} represents the
leading (zeroth) order term in an expansion in the inverse
dimensionless conductance $1/g$, terms of the next order may, under
certain (and quite typical) circumstances, produce a contribution
which can compete with, or even dominate, the contribution from $S_V
[Q_0]$.  The presence of such terms is best understood as resulting
from the spatial deformation of the zero mode induced by the spatially
inhomogeneous potential $\widehat{V} (\vect{r})$.

Formally such terms could have been accounted for by seeking spatially
inhomogeneous solutions of Eq.~\eqref{eq:saddl} at finite values of
$\widehat{V}$. A simpler computational scheme, however, is made
possible by the fact that, to the leading order, $\widehat{V}$ couples
\emph{linearly} to the inhomogeneous modes of $Q$. Thus, the
contribution of the inhomogeneous saddle point is equivalent to the
result of a Gaussian integration around the \emph{homogeneous} saddle
point.

Employing a scheme which was originally introdu\-ced by Kravtsov and
Mirlin~\cite{kravtsov_mirlin} to explore the impact of higher mode
corrections on the universal non-perturbative random matrix
correlations in a disordered metallic sample, let us parametrize the
variation of $Q(\vect{r})$ on the non-linear manifold by setting
\begin{equation*}
  Q(\vect{r}) = T_0 e^{W(\vect{r})/2} \Lambda e^{-W(\vect{r})/2}
  T_0^{-1} \; ,
\end{equation*}
where the $W(\vect{r})$ is constrained by the condition that its
anticommutator with $\Lambda$ vanishes, $\{W(\vect{r}) , \Lambda
\}=0$. To avoid counting the zero mode contribution twice, the
generators of the non-uniform transverse fluctuations $W(\vect{r})$
are subject to the additional constraint $\int d\vect{r}\; W(\vect{r})
= 0$. The spatially uniform rotations $T_0$ parametrize the zero mode
as $Q_0=T_0 \Lambda T_0^{-1}$.  The effective zero-mode action is
obtained by integrating over $W$.

An expansion of the action in the powers of $W$ generates the
hierarchy of non-universal corrections to the zero-mode action
organized as a power series in the inverse dimensionless conductance
$1/g$. Since our interest is in establishing the leading contribution
to the zero-mode action rather than the investigation of $1/g$
corrections to the leading result, it is sufficient to keep only the
linear terms in the expansion of the action in $W$.

Expanding the action up to the linear order in $W$, we obtain $S[Q]
\simeq S_0[Q_0] +S_V[Q_0] + S'[W,Q_0]$, with
\begin{multline}
  S' [W,Q_0] = -\Frac{\pi\bar{\nu}}{8} \int d\vect{r}\; \str \left[- 
  D(\nabla W)^2\right]\\ 
  +\Frac{i}{2} \int d\vect{r} \; \str \left[\Lambda T_0^{-1}
  \langle\vect{r}| \widehat{\mathcal{T}} |\vect{r} \rangle T_0
  W(\vect{r})\right] \; .
\label{eq:wexpo}
\end{multline}
Since we have chosen to employ the real space representation of the
generators $W$ of the non-uniform fluctuations, Eq.~\eqref{eq:wexpo}
involves real space matrix elements of the operator
$\widehat{\mathcal{R}}$, necessitating a switch from the momentum
(scattering channels) representation employed in Eq.~\eqref{eq:svfir}.
 
In the case of local perturbations, integrating over $W$ leads to a
contribution which is simply a $1/g$ correction to $S_0[Q_0]$, and
thus can be ignored in the present study. It is worth noting, however,
that the presence of $1/g$ correction to the \emph{parametric}
correlation functions stands in marked contrast to the non-parametric
case where the leading corrections start at the $1/g^2$
order~\cite{kravtsov_mirlin}.

Concentrating for the moment on the global perturbations, we note
that, as discussed above, local values of a global $V$ are necessarily
small. Therefore, it is sufficient to approximate
$\widehat{\mathcal{T}} \simeq P^{1/2} \widehat{V} P^{1/2}$.  The
non-universal contribution to the zero-mode action $S_W[Q_0]$ is
defined as
\begin{equation*}
  S_W[Q_0] = -\ln \langle e^{-S'} \rangle_W\; .
\end{equation*}
Utilizing the contraction rule~\cite{kravtsov_mirlin,fyodorov_mirlin}
\begin{multline*}
  \langle \str [A(\vect{r}_1) W (\vect{r}_1)] \str [B(\vect{r}_2) W
  (\vect{r}_2)] \rangle \\ 
  = 2 \Pi (\vect{r}_1 - \vect{r}_2) \str [A(\vect{r}_1) B(\vect{r}_2)
  - \Lambda A(\vect{r}_1) \Lambda B(\vect{r}_2)\\ 
  - A(\vect{r}_1) L C^{\mathsf{T}} B^{\mathsf{T}} (\vect{r}_2) C L +
  A(\vect{r}_1) \Lambda L C^{\mathsf{T}} B^{\mathsf{T}} (\vect{r}_2) C
  L \Lambda] \; ,
\end{multline*}
where $A$ and $B$ are arbitrary supermatrices, and making use of the
identity $T_0 = L C^{\mathsf{T}} {T_0^{-1}}^{\mathsf{T}} C L$, one
obtains
\begin{multline}
  S_W [Q_0]
  = \Frac{(\pi \bar{\nu})^2}{2} \int d\vect{r} d\vect{r}'\;
  \Pi (\vect{r} - \vect{r}') \str \left[\widehat{V}(\vect{r})Q_0 
  \widehat{V} (\vect{r}')Q_0\right]\\
  = \Frac{1}{g} \Frac{(\pi \bar{\nu})^2}{2 \pi} \sum_{\vect{q} \ne 0}
  \Frac{1}{(\vect{q} L)^2} \str\left(\widehat{V}_{\vect{q}} Q_0
  \widehat{V}_{-\vect{q}} Q_0\right) \; ,
\label{eq:compa}
\end{multline}
where the diffusion propagator is defined as
\begin{equation}
  \Pi (\vect{r}) = \Frac{1}{\pi g} \sum_{\vect{q} \ne 0} \Frac{e^{i
  \vect{q} \cdot \vect{r}}}{(\vect{q} L)^2} \; .
\label{eq:propi}
\end{equation}
Here, as above, $g=E_c/\bar{\Delta}$ represents the dimensionless
conductance of the disordered system. The sums in
Eqs.~\eqref{eq:compa} and~\eqref{eq:propi} are restricted to
$|\vect{q}| < 1/\ell$.

The sum 
\begin{equation}
S[Q_0] = S_0[Q_0] + S_V[Q_0] + S_W[Q_0]
\end{equation}
 represents the total
zero-mode action describing parametric correlations induced by an
arbitrary external perturbation. $S_W[Q_0]$ involves only the global
part of $V$, while $S_V[Q_0]$ contains contributions from both global
and local parts of a generic $V$.  Any corrections to this expression
involve terms of higher orders in $g^{-1}$.

We are now in position to analyze the relative importance of various
terms in $S[Q_0]$. Concentrating first on the case of global
perturbations, we have to compare Eq.~\eqref{eq:compa} to the term
coming from the expansion of the action $S_V [Q_0]$ to quadratic order
in the $\widehat{V}(\vect{r})$. (We assume that $\int d\vect{r}\;
V_i(\vect{r})=0$ for all $i$, since a nonzero value of the integral
can be accommodated by a simple shift of the corresponding frequency
$\varepsilon_i$.) The corresponding contribution is (setting $\mathrm{Re}
\langle G^{\smc{r}}\rangle = 0$ for simplicity)
%
\begin{equation}
  S^{(2)}_{V} [Q_0] = \Frac{(\pi \bar{\nu})^2}{4} \int d\vect{r}
  d\vect{r}'  f_d^2 (\vect{r} - \vect{r}')
  \str\left[\widehat{V}(\vect{r})Q_0 \widehat{V}(\vect{r}')Q_0\right]
  \; .
\label{eq:glvsl}
\end{equation}
%
If $\widehat{V}(\vect{r})$ varies slowly on the length scale of the
mean free path, making use of the identity $\int d\vect{r}\;
f_d^2(\vect{r})= \tau/\pi \bar{\nu}$, one obtains
\begin{equation}
\label{eq:sv2}
  S^{(2)}_{V}[Q_0] = \tau \bar{\Delta} \Frac{(\pi \bar{\nu})^2}{4 \pi} 
  \sum_{\vect{q} \ne 0}^{1/\ell} \str \left(\widehat{V}_{\vect{q}} Q_0
  \widehat{V}_{-\vect{q}}Q_0\right) \; .
\end{equation}
Then, using $1/g \equiv \bar{\Delta} \tau d (L/\ell)^2 \gg
\bar{\Delta} \tau$, we see that in this limit the contribution from
$S_W$ dominates over $S^{(2)}_V$ since each term in the sum over
$\vect{q}$ in $S_W$ is larger than the corresponding term in
$S^{(2)}_V$ by a factor $1/\vect{q}^2\ell^2$. The latter is large over
the whole range of the summation over $\vect{q}$.  Taken together with
the contribution $S_0[Q_0]$, in the zero-dimensional limit
$\varepsilon < E_c$ these results recover the standard universal
parametric correlation functions reported in the
literature~\cite{altshuler_simons,simons_altshuler}.
Specifically, in the simplest case of the two-point correlation
function $R_{11}$, evaluation of Eq.\ (\ref{eq:sv2}) leads to the
following value of the coefficient $C(0)$ in~\eqref{eq:xdefi}: 
%
\begin{equation}
  C(0) = 8 \bar{\nu}^2 \int d\vect{r} d\vect{r}'\; V(\vect{r}) \Pi
  (\vect{r} - \vect{r}') V(\vect{r}) \; .
\end{equation}

Now, by contrast, let us consider a potential $\widehat{V} (\vect{r})$
which has a structure at scales shorter than the mean free path. The
summation in~\eqref{eq:compa} still extends only up to $q \sim
1/\ell$, while $S^{(2)}_V$ includes equally all harmonics of
$V$. Depending on the spectral composition of $\widehat{V}$, either of
these terms may be the dominant one. Crucially, however, both $S_W$
and $S^{(2)}_V$ have identical \emph{functional forms}. Again using
the case of the two-point correlation function as a representative
example, we find that the contribution from fast modes of $V$ does not
affect either the form of Eq.~\eqref{eq:sigma} or the validity of
rescaling~\eqref{eq:xdefi}. Its only effect is to redefine the coefficient
$C(0)$, or, equivalently, $x_0^2$, as described by Eq.~\eqref{eq:addx0}.

\subsection{Local Perturbations}
\label{sec:genra}
The structure of the formalism developed above is quite general,
allowing definitive statements about universality of arbitrary
many-point correlation functions to be inferred. However, its utility
as a generating functional is at present largely limited to two-point
averages: The parametrization of supermatrices $Q_0$ of high rank
presents significant technical difficulties. Deferring the discussion
of universality of many-point correlation functions to the concluding
section of the paper, let us explore the particular case of the
two-point parametric correlation function of the
DoS~\eqref{eq:r11de}. We set $\widehat{V} = \diag (V_1,V_2)$ instead
of the more restrictive notation $\widehat{V} = \diag (0,V)$ employed in
Eq.~\eqref{eq:r11de}, in order to better illustrate the universality of
the results.

%
As discussed above, in the case of local perturbations the dominant
contribution to the action comes from $S_V[Q_0]$. Remarkably, the
supertrace in $S_V[Q_0]$ can be evaluated explicitly employing
Efetov's parametrization~\cite{efetov}:
\begin{equation}
  Q_0 = \begin{pmatrix} 
  u & 0\\ 
  0 & v
  \end{pmatrix}_{\smc{ra}} \begin{pmatrix}
  \cos \hat{\theta} & i \sin \hat{\theta} \\ 
  -i \sin \hat{\theta} & -\cos \hat{\theta}
  \end{pmatrix}_{\smc{ra}} \begin{pmatrix}
  u^{-1} & 0\\ 
  0 & v^{-1} \end{pmatrix}_{\smc{ra}} \; ,
\label{eq:param}
\end{equation}
where $u$ and $v$ are unitary $4 \times 4$ supermatrices and
$\hat{\theta}$ are matrices of commuting variables
\begin{align}
  \hat{\theta}_{\smc{bf}}^{11} &= \begin{pmatrix} 
  i\theta_1 & i\theta_2\\ 
  i\theta_2 & i\theta_1 \end{pmatrix}_{\smc{tr}} &
  \hat{\theta}_{\smc{bf}}^{22} &= \begin{pmatrix}
  \theta & 0\\ 
  0 & \theta \end{pmatrix}_{\smc{tr}} \; ,
\label{eq:theta}
\end{align}
with $0 \le \theta \le \pi$ and $0 \le \theta_{1,2} < \infty$. The
explicit parametrization of $u$ and $v$ is not needed here because
they commute with $\widehat{V}$. For systems belonging to the unitary
symmetry class (i.e. where the time-reversal invariance is lifted),
$\theta_2 = 0$.

Using the parametrization~\eqref{eq:param} and~\eqref{eq:theta},
$S_V[Q_0]$ can be compactly rewritten as
\begin{equation}
  S_V[Q_0]=\frac{1}{2}\str\ln \left(\openone +
  i\widehat{\mathcal{R}}\sigma_3^{\smc{ra}}
  e^{i\hat{\theta}\sigma_1^{\smc{ra}}}\right)\; ,
\label{eq:step1}
\end{equation}
where for convenience we have dropped the subscript $P$ from the
diagonal matrix $\widehat{\mathcal{R}} = \diag
(\mathcal{R}_1,\mathcal{R}_2)$. It is also convenient to rewrite this
expression in terms of the scattering matrices $\mathcal{S}_{1,2}$
which are related to the reactance matrices via
\begin{equation*}
  \mathcal{R}_{1,2} = i\left(\mathcal{S}_{1,2} - \openone\right)
  \left(\mathcal{S}_{1,2} + \openone\right)^{-1}\; .
\end{equation*}

In terms of 
\begin{equation}
  \widehat{\mathcal{S}} = \diag
  (\mathcal{S}_1,\mathcal{S}^{\dagger}_2)\; , 
\label{eq:smatr}
\end{equation}
Eq.~\eqref{eq:step1} takes the form
\begin{equation*}
  S_V[Q_0]= \Frac{1}{2} \str\ln \left[\openone +
  e^{i\hat{\theta}\sigma_1^{\smc{ra}}} + \widehat{\mathcal{S}}
  \left(\openone - e^{i\hat{\theta} \sigma_1^{\smc{ra}}}\right)\right]
  \; ,
\end{equation*}
where we have used $\str \ln (\widehat{\mathcal{S}} +
\openone)=0$. Note that the appearance of $\mathcal{S}^{\dagger}$
in~\eqref{eq:smatr} is a consequence of the factor
$\sigma_3^{\smc{ra}}$ in~\eqref{eq:step1} and of the identity
$\mathcal{S} (-\mathcal{R}) = \mathcal{S}^{\dag}
(\mathcal{R})$. Eq.~\eqref{eq:step1} can be rewritten as
\begin{equation}
  \Frac{1}{2} \str\ln \left(\openone + e^{i\hat{\theta}
    \sigma_1^{\smc{ra}}}\right) + \Frac{1}{2} \str\ln\left(\openone -
    i\widehat{\mathcal{S}} \sigma_1^{\smc{ra}} \tan
    \frac{\hat{\theta}}{2}\right)\; .
\label{eq:step2}
\end{equation} 
Since $\widehat{\mathcal{S}}$ is diagonal in the $\smc{ra}$ space, and
$\sigma_1^{\smc{ra}}$ is off-diagonal, only even order terms in the
series  expansion of the second logarithm in Eq.~\eqref{eq:step2}
gives a non-vanishing contribution to the supertrace. As a result,
this term can be rewritten as
\begin{equation}
  \Frac{1}{4} \str\ln \left(\openone + \widehat{\mathcal{S}}_{12}
  \tan^2\frac{\hat{\theta}}{2}\right)\; ,
\label{eq:step3}
\end{equation} 
where $\widehat{\mathcal{S}}_{12} = \diag (\mathcal{S}_1
\mathcal{S}_2^{\dag} , \mathcal{S}_2^{\dag} \mathcal{S}_1)$. A similar
transformation brings the first logarithm in Eq.~\eqref{eq:step2} to
the form
\begin{equation}
  \Frac{1}{4} \str\ln\left(\openone + \cos\hat{\theta}\right)\; .
\label{eq:step4}
\end{equation} 
Combining equations~\eqref{eq:step3} and~\eqref{eq:step4} we find 
\begin{gather}
\label{eq:step5}
  S_V[Q_0] = \Frac{1}{4} \str \ln\left(\openone +
  i\widehat{\mathcal{R}}_{12} \cos\hat{\theta}\right)\; ,
\intertext{where}
  \mathcal{R}_{12} = i\left(\mathcal{S}^{\dag}_2 \mathcal{S}_1 -
  \openone\right) \left(\mathcal{S}^{\dag}_2 \mathcal{S}_1 +
  \openone\right)^{-1}\; , 
\nonumber
\end{gather}
and, utilizing the cyclic invariance of the trace,
$\widehat{\mathcal{R}}_{12} = \diag (\mathcal{R}_{12} ,
\mathcal{R}_{12})$.

Performing the trace over the $\smc{bf}$, $\smc{ra}$ and $\smc{tr}$
indices in~\eqref{eq:step5}, and again using the cyclic invariance of the
trace, one obtains the effective action:
\begin{equation}
  S_V[Q_0] = \Frac{1}{2} \tr \ln \left[\Frac{\openone + 2i
  \mathcal{R}_{12} \lambda_1 \lambda_2 - \mathcal{R}_{12}^2
  (\lambda_1^2 + \lambda_2^2 -1)}{(\openone + i \mathcal{R}_{12}
  \lambda)^2}\right] \; ,
\label{eq:intge}
\end{equation}
where $\lambda_{1,2} = \cosh \theta_{1,2}$ and $\lambda = \cos
\theta$. In the unitary ensemble, the corresponding action can be
inferred from Eq.~\eqref{eq:intge} simply by setting
$\lambda_2=1$. $S_V[Q_0]$ is thus easily seen to coincide with
$\sigma_{\text{loc}}$, provided $x_a$ are identified with the eigenvalues of
$\mathcal{R}_{12}$.

Finally, differentiating $\langle \mathcal{Z} [j]\rangle$
\eqref{eq:sourc} with respect to $j$ and $j^\dag$, generates the source
term for the two-point correlation function of the form $(\str Q_0
\sigma_3^{\smc{ra}} \sigma_3^{\smc{bf}})^2$. Integrating over the
degrees of freedom contained in the matrices $u$ and $v$, one obtains
the general two-point correlator of DoS for the orthogonal and unitary
symmetry class, Eqs.~\eqref{eq:ranko} and~\eqref{eq:sline}.

\section{Universality and Connection to RMT}
\label{sec:couni} 
%
\subsection{Universality}
\label{sec:unive} 
Although a specific model of $H_0$ and $H_{\text{dis}}$ was used in
the calculation, the results are valid for any disordered system whose
spectral statistics exhibit the Wigner-Dyson phenomenology, since the
only crucial `ingredients' in the derivation were the existence of a
unique zero mode in the NL$\sigma$M description, and a clear
distinction between massive and soft modes. The results are also valid
for generic chaotic systems, although a special consideration may be
needed to properly take into account the contribution from the
Lyapunov region to the non-universal
terms~\cite{aleiner_larkin,kogan_efetov}.

The calculation in Section~\ref{sec:genra} underscores the
universality of the two-point parametric correlation functions in the
case of local perturbations by demonstrating that the phenomenological
parameters $y_a$ are the eigenphases of the scattering matrix
$\mathcal{S}_{12}\equiv \mathcal{S}_2^{\dag}\mathcal{S}_1$. The
importance of this result lies in the fact that $\mathcal{S}_{12}$
describes the scattering off the potential $V_2 - V_1$ when the
unperturbed Hamiltonian is $H_0 + V_1 + H_{\text{dis}}$. Thus, all
dependence on the `reference' Hamiltonian $H_0$ is excluded from the
result: parametric correlations between \emph{any} $H_1$ and $H_2$ are
parametrized by the eigenphases of the scattering matrix off the
potential $H_2 - H_1$ with $H_1$ playing the role of the unperturbed
Hamiltonian (or, equivalently, scattering off $H_1 - H_2$ with $H_2$
as the unperturbed background).

In the two-point case, as we have just shown, the correlation
functions between \emph{two} Hamiltonians $H + V_1$ and $H + V_2$
depend on a \emph{single} reactance matrix $\mathcal{R}_{12}$. At the
same time, the action~\eqref{eq:svfir} suitable for the calculation of
a generic $p$-point correlation function involving $n$ different
values of the perturbing potential $V_i$ apparently depends on the
full complement of $n$ reactance matrices $\mathcal{R}_i$.  On the
other hand, a straightforward generalization of the notion of
translational invariance in the space of Hamiltonians from two-point
functions discussed in the Introduction to the general case
immediately leads to the conclusion that such correlation functions
should depend only on the $n-1$ `mutual' reactance matrices which can
be chosen as, \emph{e.g.}, $\mathcal{R}_{1i}$, $i=2,\dots,n$.

In the absence of an explicit parametrization of the $8p\times 8p$
$Q$-matrices, it is not possible to perform a calculation analogous to
the one in Section~\ref{sec:genra} to demonstrate that this is indeed
the case. However, such a calculation is, in fact, not necessary. The
translational invariance can be inferred instead from the analysis of
the massive modes in Section~\ref{sec:massi}. Indeed, it was shown
there that locally strong perturbations do not lead to significant
coupling between massive and soft modes.  Consequently, without loss
of generality one can absorb $V_1$ into $H_0$ and redefine the
remaining potentials as $V_i \to V_i - V_1$. The corresponding
reactance matrices would be automatically redefined as
$\mathcal{R}_1\to 0$, $\mathcal{R}_i \to\mathcal{R}_{1i}$.
%
%
\subsection{Connection to RMT analysis}
\label{sec:rmtco} 
The $n=2$ case for arbitrary $p$ (and $\beta=2$) has been studied
recently using RMT techniques~\cite{SMS,SS2}. In order to extend the
RMT results to generic disordered/chaotic systems a phenomenological
\emph{ansatz} based on Berry's conjecture~\cite{berry} was employed
in~\cite{SMS}. The present analysis affords an opportunity to achieve
such an extension in a more rigorous way.

We begin by noting that in the $p=2$ case the \emph{structure} of the
2-point parametric correlation functions obtained above for the
disordered systems with broken time-reversal invariance is
\emph{identical} to the structure of the results obtained
in~\cite{SMS,SS2}. Indeed, according to Refs.~\cite{SMS,SS2} the
two-point parametric correlation function in random matrix ensembles
of unitary symmetry is
\begin{equation*}
  R_{11} (\Omega) = 1 - \left\{\hat{\mathcal{D}}^{-1}
  \left[k(s)-\delta(s)\right]\right\} \left[\hat{\mathcal{D}}
  k(s)\right] \; .
\end{equation*}
Here, as before, $s=\Omega/\bar{\Delta}$, the differential operator
$\hat{\mathcal{D}}$ is defined as
\begin{equation*}
  \hat{\mathcal{D}} = \det\left(\openone
  - \mathfrak{R}_{\smc{rmt}} \Frac{d}{d s}\right) \; ,
\end{equation*}
the function $k (s)$ has been defined in~\eqref{eq:kfunc} and
$\mathfrak{R}_{\smc{rmt}}$ is the random matrix version of the
reactance matrix,
\begin{equation*}
  \mathfrak{R}_{\smc{rmt}} = \frac{V}{\openone - \mathrm{Re} \langle
  G^{\smc{r}} \rangle V}\; .
\end{equation*}
In the standard random matrix ensembles the real part of the average
Green function is a diagonal matrix, thus simplifying the structure of
the reactance matrix. Using the Fourier transforms
\begin{gather*}
  k(s) = \int_{-1}^{1}\Frac{d\lambda}{2} e^{i\pi \lambda s}
\intertext{and}
  k(s) - \delta(s) = -\left(\int_{-\infty}^{-1} +
  \int_{1}^{\infty}\right) \Frac{d\lambda_1}{2} e^{i\pi \lambda_1 s}\;
  ,
\end{gather*}
we immediately recover Eqs.~\eqref{eq:sline} and~\eqref{eq:sgene},
where $x_a$ are the eigenvalues of $\mathfrak{R}$. It is intriguing
that the supersymmetric structure involving the pair of compact
($\lambda$) and non-compact ($\lambda_1$) variables is reproduced
under the guise of the dual pair
$(\hat{\mathcal{D}},\hat{\mathcal{D}}^{-1})$ as a result of the RMT
analysis based solely on the method of orthogonal polynomials
\cite{SMS,SS2}.

It was shown in~\cite{SMS,SS2} that in unitary random matrix ensembles
the $p$-point parametric correlation function has 
the form
\begin{widetext}
\begin{multline}
  R_{p_1 p_2} (\{\varepsilon_i\}_{i=1}^p) = \bar{\Delta}^p \langle
  \prod_{i=1}^{p_1} \tr \delta(\varepsilon_i - H) \prod_{j'= p_1 +
  1}^{p} \tr \delta(\varepsilon_{j'} - H - V) \rangle\\
  = \det \begin{pmatrix}
  k (s_i - s_j) 
  & \hat{\mathcal{D}}^{-1} \left[k(s_i - s_{j'}) -
  \delta(s_i - s_{j'})\right] \\
  \hat{\mathcal{D}} k(s_{i'} - s_j) & k (s_{i'} - s_{j'}) 
  \end{pmatrix} \; ,
\label{eq:nmcor}
\end{multline}
\end{widetext}
where $s_i = \varepsilon_i /\bar{\Delta}$, the indices $i$ and $j$ run
over the range $1,\dots, p_1$, and $i'$ and $j'$ are in the range
$p_1+1,\dots, p$. All energies $\varepsilon_i$ inside the subsets
$[1,p_1]$ and $[p_1+1,p]$ are assumed to be different, which
corresponds to neglecting the $\delta$-function terms describing
self-correlations of levels. The generalization of
Eq.~\eqref{eq:nmcor} to the case when some energies coincide can be
found in~\cite{SMS,SS2}. We now note that, in complement to the
orthogonal polynomial method used in Refs.~\cite{SMS,SS2}, the
parametric correlation functions in random matrix ensembles can be
alternatively studied using the non-linear sigma model approach. The
resulting sigma model action has the form of Eq.~\eqref{eq:sinte2}
parametrized by $\mathfrak{R}_{\smc{rmt}}$. Since they are described
by the same action, the correlation functions have the same
\emph{functional form} irrespective of whether the averaging is
performed over $H$ drawn from an invariant distribution or over $H =
H_0 + H_{\text{dis}}\equiv \hat{\xi}_{\hat{\vect{p}}} + U(\vect{r})$
where the distribution of $U(\vect{r})$ is described by
Eq.~\eqref{eq:varia}. It follows that \emph{the $p$-point correlation
functions in generic disordered/chaotic systems of unitary symmetry
are given by the universal Eq.~\eqref{eq:nmcor} with the operator
$\hat{\mathcal{D}}$ parametrized by the corresponding reactance matrix
$\mathcal{R}$.} This conclusion about universality and parametrization
extends also to the level-number-dependent correlation functions
studied in~\cite{SS1,SS2} since the latter are based on
Eq.~\eqref{eq:nmcor}.

\begin{acknowledgments}
  One of us (FMM) would like to acknowledge the financial support in
  part of Cofinanziamento MIUR (prot. 2002027798) and TCM group,
  and in part of EPSRC (GR/R95951).
\end{acknowledgments}


\end{document}